\documentclass[]{interact}

\usepackage{epstopdf}
\usepackage{comment}
\usepackage[longnamesfirst,sort]{natbib}
\bibpunct[, ]{(}{)}{;}{a}{,}{,}

\usepackage{multirow}
\usepackage{hyperref}
\usepackage{amsfonts}
\usepackage{wasysym}
\usepackage{color}
\usepackage{graphicx}
\usepackage[none]{hyphenat}
\usepackage[utf8]{inputenc}
\usepackage[english]{babel}
\usepackage{amsmath}
\usepackage[euler]{textgreek}
\usepackage{dsfont}
\usepackage{afterpage}
\usepackage{booktabs}
\usepackage{caption}
\usepackage{subcaption}
\usepackage{float}
\usepackage{pdflscape}
\usepackage{graphicx}
\usepackage{ragged2e}
\usepackage{amssymb}
\usepackage{amsthm}
\usepackage{natbib}
\usepackage{xcolor}
\usepackage{lipsum}
\usepackage{setspace}
\usepackage{parskip}
\usepackage{geometry}
\usepackage{titlesec}
\usepackage{fancyhdr}
\usepackage{enumerate}
\usepackage{centernot}
\usepackage{listings}
\usepackage{etoolbox}
\usepackage{mathtools}
\usepackage{algpseudocode,algorithm}
\usepackage{times}
\usepackage{pdfpages}
\usepackage{float}
\usepackage{tikz}
\usepackage{enumitem}

\newcommand{\E}{\mbox{\sf E}} 
\newcommand{\BC}{\begin{center}}
\newcommand{\EC}{\end{center}}
\newcommand{\BE}{\begin{equation}}
\newcommand{\EE}{\end{equation}}
\newcommand{\BEA}{\begin{eqnarray}}
\newcommand{\EEA}{\end{eqnarray}}
\newcommand{\BEAS}{\begin{eqnarray*}}
\newcommand{\EEAS}{\end{eqnarray*}}

\newcommand{\dg}[1]{{\color{red}{#1}}}

\begin{document}

\title{Cost-Efficient Fixed-Width Confidence Intervals for the Difference of Two Bernoulli Proportions}

\author{$^\star$ \and $^\star$}

\author{
\name{Ignacio Erazo\textsuperscript{a}, \; David Goldsman\textsuperscript{a}, \; Yajun Mei\textsuperscript{a} \thanks{CONTACT David Goldsman. Email: sman@gatech.edu}}
\affil{\textsuperscript{a}H. Milton Stewart School of Industrial and Systems Engineering,
Georgia Institute of Technology, Atlanta, GA 30332-0205, U.S.A.}
}

\bibliographystyle{plain}

\maketitle

\begin{abstract}

We study properties of confidence intervals (CIs) for the difference of two Bernoulli distributions' success parameters, $p_x - p_y$, in the case where the goal is to obtain a CI of a given half-width while minimizing sampling costs 
when the observation costs may be different between the two distributions.  
We propose three different methods for constructing fixed-width CIs: (i) 
a two-stage sampling procedure,  (ii) a sequential method that carries out sampling in batches, and (iii) an $\ell$-stage ``look-ahead'' procedure. We use Monte Carlo simulation to show that, under diverse success probability and observation cost scenarios, our proposed algorithms obtain significant cost savings versus their baseline counterparts (up to 50\% for the two-stage procedure, up to 15\% for the sequential methods). Furthermore, for the battery of scenarios under study, our sequential-batches and $\ell$-stage ``look-ahead'' procedures approximately obtain the nominal coverage while also meeting the desired width requirement. In addition, our sequential-batching method  turned out to be more efficient than the ``look-ahead'' method from a computational standpoint, with average running times at least an order-of-magnitude faster over all the scenarios tested. We illustrate the cost-efficacy of our new procedures on a case study involving a comparison between the effectiveness of generic and brand-name drugs. 
\end{abstract}


\section{Introduction and Motivation}
\label{sec:intro}

This article proposes various efficient procedures to obtain confidence intervals (CIs) for the difference $p_x-p_y$ of two Bernoulli distributions' success parameters, $p_x$ and $p_y$. Our goal is to obtain a CI of a pre-specified width while minimizing sampling costs, where the observation costs may be different between the two distributions. This paper complements \cite{erazo21}, which assumed that sampling costs are equal and so instead simply sought to minimize the number of observations taken. At the onset, we present several examples to motivate our interest in addressing cost considerations. 
\begin{itemize}

\item An obvious application is to compare the likelihood of a good outcome when using a inexpensive generic drug to the likelihood of a good outcome using the corresponding expensive brand-name drug. The drug cost difference comes into play as we seek to obtain valid statistical conclusions in a parsimonious way.   And the consequences of the exercise are relevant in the real world because if no perceived difference in drug performance is found, then use of the generic version can lead to huge societal savings.

\item In fact, in many biological or biomedical studies, one is often interested in investigating statistical inference on the difference in the observed risk of events between experimental and control interventions. It stands to reason that the financial burdens and consequences are different in two groups when considering costs and benefits of implementing different interventions. For instance, we might be interested in probability that hemoglobin A1C will be at an acceptable level if (i) a patient follows a “good eating and exercise” regimen (which is relatively inexpensive to conduct) vs.\ (ii) the patient takes an expensive experimental drug, coupled with proper diet and exercise.
    
\item We might wish to compare the probability that a person will pass an automated online interview (inexpensive to carry out) with the probability that a person will pass a live interview conducted in-person (requiring more-expensive resources).
    
\item An analyst wants to undertake real-world experiments to estimate the probability that a manufacturing plant can deliver 85\% of its orders on time during a given week.  Under scenario one, the plant manufactures the orders in a just-in-time manner and maintains minimal inventory (which is relatively inexpensive).  But under scenario two, the plant tends to carry additional inventory in the hope of having sufficient stock on hand for upcoming orders (which is relatively more-expensive).

\end{itemize}

It is well known that taking numerous observations from both Bernoulli populations will eventually whittle down a CI for $p_x-p_y$ to any desired half-width $\epsilon$.  Our interest lies in achieving a small half-width $\epsilon$ requirement while (approximately) maintaining  the nominal coverage probability of $1-\alpha$, all the while doing so at the lowest cost --- which is not necessarily the same as achieving the nominal coverage and desired half-width using the fewest observations. One population's observations may simply be significantly more expensive than the other's; and therefore, we ought to consider sampling schemes giving us more ``bang for the buck''. 

In this paper, we propose a number of sampling schemes, all of which assume that we have in hand preliminary estimates of $p_x$ and $p_y$ (which may be obtained via  stage-one samples from the two populations). In particular, we consider the following sampling schemes:

\begin{itemize}
    \item A two-stage procedure, presented in Algorithm \ref{Algo:2StageMinCost};
    \item A sequential-sampling procedure carried out in batches, presented in Algorithm \ref{Algo:BatchSeqMinCost};
    \item An $\ell$-stage ``look-ahead'' procedure that also carries out the sampling in (time-saving) batches, presented in Algorithm \ref{Algo:1StageLookAhead}.
\end{itemize}
We also present a conservative benchmark procedure (Algorithm \ref{Algo:Conservative}) to compare against our two-stage sampling scheme, and a ``naive'' benchmark procedure (Algorithm \ref{Algo:NaiveSeqMinCost}) to compare against our sequential-sampling scheme. 
Under a diverse battery of practical scenarios for the values of the success probabilities and observation costs, we use Monte Carlo simulation as well as a case study involving the efficacy of generic and brand-name drugs to evaluate procedure performance; and our results indicate that:
\begin{itemize}
    \item Our procedures (approximately) obtain the nominal coverage $1-\alpha$;
    \item The procedures obtain the desired CI width, except in the case of the two-stage procedure (Algorithm \ref{Algo:2StageMinCost}) which relies on a heuristic that yields a {\em random\/} width that is correct ``on average'', yet exhibits substantial variability.
    \item The procedures do well in terms of reducing costs, especially when compared to the benchmark procedures or to methods that merely minimize the total number of observations used \citep{erazo21} --- none of which take costs into consideration.
\end{itemize} 

The rest of this paper proceeds as follows. \S\ref{sec:background} provides additional background material, including relevant notation. In \S\ref{sec:SScostmin}, we describe a two-stage procedure (Algorithm \ref{Algo:2StageMinCost}) that incorporates a cost-minimization strategy. 
\S\ref{sec:SS-group} presents a sequential procedure (Algorithm \ref{Algo:BatchSeqMinCost}) in which observations are taken in batches in order to reduce the number of sampling stages (thus saving time). \S\ref{sec:lstage} presents what we call the $\ell$-step look-ahead procedure (Algorithm \ref{Algo:1StageLookAhead}) that performs sampling in batches, while considering potential future outcomes. \S\ref{sec:results} discusses a battery of Monte Carlo evaluations of our various proposed algorithms over an array of parameter configurations; notably, we present a detailed case study comparing the efficacy and costs of generic vs.\ brand-name drugs in \S\ref{sec:studycase}.  We establish that our proposed procedures perform as advertised, often resulting in substantial savings over their baseline counterparts and their counterparts that attempt to  minimize the total number of observations taken without explicitly considering costs. We summarize our work and provide conclusions in \S\ref{sec:conclusions}. 

\section{Background}
\label{sec:background}

\S\ref{subsec:notation}
 spells out the general problem at hand and establishes the necessary notation.  We also provide a brief summary of relevant work in the area.  \S\ref{subsec:conservative}  puts forth a simple straw man conservative procedure that we can use for comparison purposes later on.

\subsection{Some Notation and Literature}
\label{subsec:notation}


Going forward, we assume that we have at our disposal a stream  $X_1,X_2,\ldots$ of independent and identically distributed (iid) Bern($p_x$) random variables from population $X$ as well as an iid Bern($p_y$) stream $Y_1,Y_2,\ldots$ from population $Y$; and moreover, the $X$'s and $Y$'s are independent of each other. This independent-samples set-up is common in practice, and is especially useful in the context of computer simulations, where we have the ability to control the random number streams driving the simulations so as to produce independent replications of any experiments. 

As a baseline, we consider the following classical Wald CI for $p_x-p_y$, which is familiar from any elementary statistics textbook \citep{hines2003}, 
\BE
p_x - p_y \; \in \;   \bar{X} - \bar{Y} \pm H \; \equiv \; \bar{X} - \bar{Y} \pm z_{\alpha/2}\sqrt{\frac{\bar{X}(1-\bar{X})}{m_x} + \frac{\bar{Y}(1-\bar{Y})}{m_y}},
\label{eq:original-ci}
\EE
where the respective fixed sample sizes are $m_x$ and $m_y$, the corresponding sample means are given by $\bar{X} \equiv \sum_{i=1}^{m_x}X_i/m_x$ and $\bar{Y} \equiv \sum_{i=1}^{m_y}Y_i/m_y$,  $z_{\alpha/2}$ is the  $(1- \frac{\alpha}{2})$ quantile of the standard normal distribution, and $H$ denotes the half-width defined by the right-hand-side expression. 
As described in \cite{erazo21}, variants of and alternatives to the one-dimensional Bern($p$) version of the Wald CI are surveyed in  e.g.,  \cite{ac98}, \cite{bcd01, bcd02}, \cite{f10}, and \cite{turner13}.

Although the Wald CI suffers from poor small-sample
properties (e.g., the coverage probability can substantially deviate from the nominal  $1-\alpha$  for small $m_x$ and $m_y$), 
the sample sizes we deal with  in  the current paper will be substantial enough so that our proposed sequential procedures typically obtain the nominal coverage, at least approximately --- a claim that we validate via simulation.

In this paper, we take observation costs into consideration --- all observations are not created equal.  To this end, let $c_x > 0$ and $c_y > 0$  denote the costs of taking a single observation from populations $X$ and $Y$, respectively.
Our job is to deliver a cost-efficient CI on the difference $p_x-p_y$ of the form $\hat{p}_x - \hat{p}_y \pm H$ that in some sense minimizes the expected cost $\E[c_x N_x + c_y N_y]$ subject to: (i) a half-width constraint
$H\leq \epsilon$ for some desired value $\epsilon > 0$, while (ii) approximately maintaining the nominal coverage, $\Pr\!\big(
 p_x - p_y \in \hat{p}_x - \hat{p}_y \pm H \big) \ge 1 - \alpha$, where, for now,  $\hat{p}_x$ and $\hat{p}_y$ denote generic estimators of $p_x$ and $p_y$, and $N_x$ and $N_y$ denote the ultimate sample sizes from the two populations.
Such a half-width requirement typically requires more than one stage of sampling to be undertaken. 
With regard to fixed-width sequential methods for the one-dimensional Bern($p$) case, there is a rich literature; see, for instance,  
\cite{armitage1958},
\cite{k98}, 
\cite{robbins74},
\cite{gym17}, 
and \cite{nitiszacks07}, just to name a few.


\subsection{Conservative Approach}
\label{subsec:conservative} 

We start with a very simple conservative procedure.
At the beginning of the process of constructing a CI for $p_x-p_y$ with the desired half-width $H\leq \epsilon$, the decision maker may have on hand preliminary estimates $\hat{p}_x, \hat{p}_y $ for $p_x,  p_y $ (perhaps obtained from historical data, expert opinion, manufacturer specifications, etc.). These estimates may be quite accurate; however, it is not possible for a single stage of sampling to guarantee the desired half-width requirement $H \le \epsilon$ for all possible sample paths unless a very conservative  approach is used. A common way to address this issue is by noting that the function $p(1-p)$ is maximized at $p= 0.5$, and then assuming $\hat{p}_x=\hat{p}_y=0.5$ in order to calculate a deterministic ``worst-case'' sample size that will always produce the desired half-width. Under that worst-case scenario, Expression \eqref{eq:original-ci} yields
\[
H \; \le \; \frac{z_{\alpha/2}}{2}\sqrt{\frac{1}{m_x}+\frac{1}{m_y} }.
\] 
Moreover, given a total sample-size budget of $m_x+m_y=m$, we see that the right-hand side of the above expression is minimized by $m_x=m_y$, in which case
\[
    H \; \le \; z_{\alpha/2}\sqrt{\frac{1}{2m_x}} \; \leq \; \epsilon \quad \Leftrightarrow \quad
   m_x \; \geq \; \frac{ z_{\alpha/2}^2}{2\epsilon^2}, 
\]
yielding a lower bound on $m_x$ --- depending only on $\epsilon$ and $\alpha$ --- that guarantees $H \le \epsilon$.
 This approach is described in Algorithm \ref{Algo:Conservative}, and the end result is a conservative CI meeting the half-width requirement after taking $m_x=m_y$ observations from both populations $X$ and $Y$. The procedure is conservative in the sense that the half-width will likely be somewhat smaller than called for (indicating the waste of some unnecessary observations).

\begin{algorithm}\caption{Conservative Procedure}\label{Algo:Conservative}
\textbf{Notation:} $m_x, m_y$: common number of observations to take from populations $X$ and $Y$.
\begin{algorithmic}[1]
\Require Desired coverage probability ($1-\alpha$); desired half-width $\epsilon$.
\State $m_x, m_y \leftarrow \bigg\lceil\dfrac{z_{\alpha/2}^2 }{2\epsilon^2}\bigg\rceil$, where $\lceil \cdot \rceil$ denotes the ``ceiling'' (integer round-up) function
\State Execute the sampling with $m_x$ and $m_y$
\State Compute $\bar{X}$ and $\bar{Y}$
\Ensure Confidence Interval $\bar{X} - \bar{Y} \pm z_{\alpha /2} \sqrt{ \frac{\bar{X}(1-\bar{X})}{m_x} + \frac{\bar{Y}(1-\bar{Y})}{m_y} }$
\end{algorithmic}
\end{algorithm}

\section{Two-Stage Confidence Interval Cost-Minimization Procedure}
\label{sec:SScostmin}

Henceforth, we assume that we have at our disposal preliminary stage-one estimates $\hat{p}_x$ and $\hat{p}_y$ of the respective parameters $p_x$ and $p_y$.  These estimates could be based on, e.g., expert opinion, prior experience, or actual first-stage samples of sizes $m_{x,1}$ and $m_{y,1}$ taken from populations $X$ and $Y$.  We would now like to conduct a second stage of sampling to obtain an ``inexpensive'' CI that also satisfies the half-width requirement --- or at least {\em aims\/} at satisfying the requirement. We also assume that each observation from population $X$ [$Y$] incurs a cost of $c_x >0$ [$c_y >0$]. For ease of exposition, we will first assume for now that $\hat{p}_x$ and $\hat{p}_y$ are derived from expert opinion, allowing us to take $m_{x,1} = m_{y,1} = 0$ as the first-stage sample sizes.

We formulate and solve the following optimization problem in order
to determine the total sample sizes $m_{x,2}$ and $m_{y,2}$ to be taken by the time stage two is completed. 
\begin{align}
    \min \ \ & m_{x,2} c_x + m_{y,2} c_y \label{LP1.1}  \\
    s.t. \ \  & z_{\alpha/2} \sqrt{\frac{\hat{p}_x(1-\hat{p}_x)}{m_{x,2}}+\frac{\hat{p}_y(1-\hat{p}_y)}{m_{y,2}}} \leq \epsilon \label{LP1.2}\\
    & m_{x,2},m_{y,2} \in \mathbb{Z}^+ .   \label{LP1.3} 
\end{align}

Constraint \eqref{LP1.2} arises from Equation \eqref{eq:original-ci} with the preliminary estimates $\hat{p}_x$ and $\hat{p}_y$ substituting for $\bar{X}$ and $\bar{Y}$, respectively. We let $\tau_x \equiv\hat{p}_x (1-\hat{p}_x)$ and $\tau_y \equiv \hat{p}_y(1-\hat{p}_y)$, so that \eqref{LP1.2} is equivalent to the condition 
\[
 \frac{\tau_x}{m_{x,2}}+\frac{\tau_y}{m_{y,2}} \; \leq \; \frac{\epsilon^2}{z_{\alpha/2}^2}.
 \]
 Moreover, by setting $\zeta_x \equiv \frac{1}{m_{x,2}}>0$ and $\zeta_y \equiv \frac{1}{m_{y,2}}>0$ and relaxing the integrality of $m_{x,2}$ and $m_{y,2}$, we obtain the following equivalent problem.
\begin{align}
    \min \ \ & \frac{c_x}{\zeta_x} + \frac{c_y}{\zeta_y} \label{LP2.1}\\
    s.t. \ \ & \zeta_x\tau_x+\zeta_y\tau_y \; \leq \; \frac{\epsilon^2}{z_{\alpha/2}^2} \label{LP2.2} \\
    & \zeta_x,\zeta_y > 0.
\end{align}

This is a convex problem, and the Karush-Kuhn-Tucker (KKT) first-order sufficient conditions for optimality \citep{KKT} ensure the existence of a Lagrange multiplier $\lambda\geq 0$ such that 
\[
\zeta_x \; = \; \sqrt{\frac{c_x}{\lambda \tau_x}}, \quad
\zeta_y \; = \; \sqrt{\frac{c_y}{\lambda \tau_y}}, \quad \mbox{and} \quad
\zeta_x\tau_x+\zeta_y\tau_y \; = \;  \frac{\epsilon^2}{z_{\alpha/2}^2}.
\]
After a little algebra involving the above three equations, we easily find that
\begin{align}
        \lambda &= \bigg(\frac{\sqrt{\tau_x c_x}+\sqrt{\tau_y c_y}}{\epsilon^2/z_{\alpha/2}^2} \bigg)^2. \label{KKT4}
\end{align}
Since $\lambda>0$,  all KKT conditions are met, and so we have an optimal solution for the convex problem. In particular, the optimal $\zeta_x$ and $\zeta_y$ are obtained by substituting the value of $\lambda$ from \eqref{KKT4}. Thus, the approximate solution for the two-stage sampling problem is obtained by setting $m_{x,2} = \lceil 1/\zeta_x \rceil$ and $m_{y,2} = \lceil 1/\zeta_y \rceil$. 

We now consider the specific case in which $\hat{p}_x$ and $\hat{p}_y$ are sample means arising from recent (first-stage) samples of sizes $m_{x,1}$ and $m_{y,1}$, respectively. How many new observations do we have to take in order to obtain our desired CI\@? Our strategy is to solve the optimal-$(m_{x,2},m_{y,2})$ problem just as before, and we see that there are two outcomes of note:

\begin{enumerate}[label=(\roman*)]
\item If we obtain $m_{x,2}\geq m_{x,1}$ and $m_{y,2} \geq m_{y,1}$, then the numbers of new observations to take in stage two will be $b_{x}=m_{x,2}-m_{x,1}$ and $b_{y}=m_{y,2}-m_{y,1}$ for each population, respectively.
\item Suppose that either $m_{x,2} < m_{x,1}$ or $m_{y,2} < m_{y,1}$ (but not both, else we would already have the desired CI)\@. Without loss of generality, we assume only that $m_{x,2}<m_{x,1}$; and then all we need to do is to solve a trivial optimization problem that re-sets $m_{x,2}=m_{x,1}$ and optimizes over $m_{y,2}$,
    \begin{align*}
        \min \ \ & m_{y,2}\\
        s.t. \ \ & z_{\alpha/2} \sqrt{\frac{\tau_x}{m_{x,1}}+\frac{\tau_y}{m_{y,2}}} \; \leq \; \epsilon.
    \end{align*}
This immediately leads to the solution in which we re-set
\[
m_{y,2} \; = \; \bigg\lceil\frac{\tau_y}{ \frac{\epsilon^2}{z_{\alpha/2}^2} - \frac{\tau_x}{m_{x,1}}} \bigg\rceil,
\]
so that it is optimal to take $b_{y}=m_{y,2}-m_{y,1}$ second-stage observations from  population $Y$.
\end{enumerate}

The two-stage procedure to obtain the CI while minimizing costs is given by Algorithm \ref{Algo:2StageMinCost}. We note that it is often the case that the two-stage procedure {\em does not achieve the length requirement\/} (see the Monte Carlo discussion in \S\ref{sec:results}). This is because the two-stage procedure only {\em aims\/} at yielding a CI of the desired length --- there is simply no chance after the second (final) stage of sampling to correct for overshooting the specified length.  This issue will be addressed in the subsequent discussion on sequential sampling.

\begin{algorithm}
\caption{Two-Stage Procedure that minimizes the total sampling cost}\label{Algo:2StageMinCost}
\textbf{Notation:} $m_{x,2},  m_{y,2}$: cumulative numbers of observations taken from populations $X$ and $Y$, respectively, by the completion of the second stage.  $\bar{X}_2, \bar{Y}_2$: overall sample means at the end of the second stage of sampling.
\begin{algorithmic}[1]
\Require initial sample sizes $m_{x,1}, m_{y,1}$; initial estimates $\hat{p}_x$ and $\hat{p}_y$; desired coverage probability $(1-\alpha)$; desired half-width $\epsilon$, and individual observation  costs $c_x>0, c_y>0$  from populations $X$ and $Y$. 
\State $\tau_x \gets \hat{p}_x(1-\hat{p}_x), \; \tau_y \gets \hat{p}_y(1-\hat{p}_y)$ \vspace{.05in}
\State $\lambda \gets \bigg(\frac{\sqrt{\tau_x c_x}+\sqrt{\tau_y c_y}}{\epsilon^2/z_{\alpha/2}^2} \bigg)^{\!2}$
\State $\zeta_x = \sqrt{\dfrac{c_x}{\lambda \tau_x}}$, \; $\zeta_y = \sqrt{\dfrac{c_y}{\lambda \tau_y}}$
\State $m_{x,2} \gets \max(\lceil 1/\zeta_x \rceil , m_{x,1} ) $, \; $m_{y,2} \gets \max(\lceil 1/\zeta_y \rceil , m_{y,1} ) $
\If{ $m_{x,2} - m_{x,1}==0$ and $m_{y,2} - m_{y,1}==0$}
    \State Do not take samples
\ElsIf{ $m_{x,2} - m_{x,1}==0$ and $m_{y,2} - m_{y,1}>0$ }
    \State Re-set $m_{y,2} \gets \bigg\lceil \dfrac{\tau_y}{\frac{\epsilon^2}{z_{\alpha/2}^2} - \frac{\tau_x}{m_{x,1}} } \bigg\rceil$
    \State  Sample $m_{y,2} - m_{y,1}$ observations from Population $Y$
\ElsIf{  $m_{x,2} - m_{x,1}>0$ and $m_{y,2} - m_{y,1}==0$ }
    \State Re-set $m_{x,2} \gets \bigg\lceil \dfrac{\tau_x}{\frac{\epsilon^2}{z_{\alpha/2}^2} - \frac{\tau_y}{m_{y,1}} } \bigg\rceil$
    \State  Sample $m_{x,2} - m_{x,1}$ observations from Population $X$
    
\Else
    \State Sample $m_{x,2} - m_{x,1}$ and $m_{y,2} - m_{y,1}$ observations from populations $X$ and $Y$, respectively
\EndIf
\State Compute $\bar{X}_2$ and $\bar{Y}_2$
\Ensure Confidence Interval $\bar{X}_2 - \bar{Y}_2 \pm z_{\alpha /2} \sqrt{ \frac{\bar{X}_2(1-\bar{X}_2)}{m_{x,2}} + \frac{\bar{Y}_2(1-\bar{Y}_2)}{m_{y,2}} }$
\end{algorithmic}
\end{algorithm}


\section{Sequential-Group Sampling Procedure}
\label{sec:SS-group}

In this section, we propose sequential-group sampling procedures. Our motivations for this approach are: (i) we must {\em guarantee\/} a CI half-width of some desired length, while approximately attaining the nominal coverage; (ii) we would like to do so at minimal cost; (iii) if time is a consideration, we might wish to conduct sequential sampling in a reasonable number of stages; and (iv) there may be a maximum sampling rate per unit time --- say, up to a batch of $B \ge 1$ observations per day, where $B=1$ indicates the fully sequential case.
With these considerations in mind, we develop two sequential CI procedures: one employs what we refer to as the \emph{naive} approach, and the other aims to minimize the total sampling cost. 

For the remainder of this section, we use the following notation. Let $m_{x,r}$ and $m_{y,r}$ denote the total number of observations taken from populations $X$ and $Y$, respectively,  up to the end of sampling epoch $r \ge 1$; let $\bar{X}_{r}$ and $\bar{Y}_{r}$ be the corresponding sample means; and let $\tau_{x,r} \equiv \bar{X}_{r}(1-\bar{X}_{r})$ and $\tau_{y,r} \equiv \bar{Y}_{r}(1-\bar{Y}_{r})$. 

We first present a naive baseline procedure (Algorithm \ref{Algo:NaiveSeqMinCost}) to carry out sequential batch sampling. Although this approach ensures the desired half-width is eventually attained, it is naive in the sense that it does not use information accumulated as sampling progresses to adapt the proportion of observations to be taken from each population during successive stages --- when $B$ is even, the procedure simply takes $B/2$ observations from both populations; and if $B$ is odd, the extra observation is assigned arbitrarily. 
\clearpage
\begin{algorithm}[h!]
\caption{Naive Sequential Batching Procedure}\label{Algo:NaiveSeqMinCost}
\textbf{Notation:} $m_{x,r}, m_{y,r}$: cumulative numbers of observations taken from populations $X$ and  $Y$, respectively, at the completion of stage $r$. $\bar{X}_r, \bar{Y}_r$: sample means at the completion of stage $r$.
\begin{algorithmic}[1]
\Require initial sample sizes $m_{x,1},  m_{y,1}$; desired coverage probability $(1-\alpha)$;  sampling capacity per stage $B$ (assume with little loss of generality that $B$ is even); and desired half-width $\epsilon$
\State $r \gets 2$, Continue $\gets$ $True$
\While{Continue == $True$}
    \State Sample $ B/2$ observations from population $X$ and $B/2$ from population $Y$
    \State $m_{x,r} \gets m_{x,r-1}+B/2$, $m_{y,r} \gets m_{y,r-1}+B/2$
    \State Compute $\bar{X}_r$ and $\bar{Y}_r$
    \State $\tau_{x,r}\gets \bar{X}_r(1-\bar{X}_r) $,  $\tau_{y,r}\gets \bar{Y}_r(1-\bar{Y}_r) $
    \State $H_r\gets z_{\alpha/2}\sqrt{ \dfrac{\tau_{x,r}}{m_{x,r}} + \dfrac{\tau_{y,r}}{m_{y,r}}}$
    \If{$H_r\leq \epsilon$}
        \State Continue $\gets$ $False$
    \Else
        \State $r \gets r+1$
    \EndIf
\EndWhile
\Ensure Confidence Interval $\bar{X}_r - \bar{Y}_r \pm H_r$
\end{algorithmic}
\end{algorithm}

Next, we propose what could be considered as our main, recommended method: a slightly more-elaborate sequential procedure (Algorithm \ref{Algo:BatchSeqMinCost}) that incorporates batching to save on the number of stages; but in addition, it makes use of the new information obtained as sampling proceeds, so as to reduce sampling cost.
Motivated by the results from \S\ref{sec:SScostmin}, this new procedure uses Algorithm \ref{Algo:2StageMinCost} iteratively as a subroutine, and at each step adapts its results so as to obtain a sequential algorithm. The new procedure is defined explicitly in Algorithm \ref{Algo:BatchSeqMinCost} and is described in a more-conceptual manner as follows.

For each iteration $r$:

\begin{enumerate}[label=(\roman*)]

\item Use the (previous) sample means $\bar{X}_{r-1}$ and $\bar{Y}_{r-1}$ to compute $\tau_{x,r-1} = \bar{X}_{r-1}(1-\bar{X}_{r-1})$ and $\tau_{y,r-1} = \bar{Y}_{r-1}(1-\bar{Y}_{r-1})$. Then compute 
\[
\lambda_r = \bigg(\dfrac{\sqrt{\tau_{x,r-1} c_x}+\sqrt{\tau_{y,r-1} c_y}}{\epsilon^2/z_{\alpha/2}^2} \bigg)^2, \quad
\zeta_{x,r} = \sqrt{\dfrac{c_x}{\lambda_r \tau_{x,r-1}}}, 
\quad \mbox{and} \quad
\zeta_{y,r} = \sqrt{\dfrac{c_y}{\lambda_r \tau_{y,r-1}}}. 
\]
These computations correspond to Steps 1, 2, and 3 of Algorithm \ref{Algo:2StageMinCost}. 
   
\item  With the values computed in (i) in hand, we use Algorithm \ref{Algo:2StageMinCost} to \emph{estimate} the number of observations that would be required (using only one extra stage) from each population in order to obtain a CI of the desired length.  This corresponds to $\max(\lceil 1/\zeta_{x,r} \rceil,m_{x,r-1}) - m_{x,r-1}$ observations from population $X$ and $\max(\lceil 1/\zeta_{y,r} \rceil,m_{y,r-1}) - m_{y,r-1}$ from population $Y$. As we only have a total sampling capacity of $B$ on stage $r$, we need to describe an efficient way to distribute the observations between the two populations. We compute the sampling proportion for population $X$, denoted by $\gamma_r$, as follows,
    \[
    \gamma_r \; \equiv \; \frac{\max(\lceil 1/\zeta_{x,r} \rceil,m_{x,r-1}) - m_{x,r-1}}{(\max(\lceil 1/\zeta_{x,r} \rceil,m_{x,r-1}) - m_{x,r-1})+(\max(\lceil 1/\zeta_{y,r} \rceil,m_{y,r-1}) - m_{y,r-1})},
    \]
    where the denominator is guaranteed to be positive (otherwise we would have already achieved the desired CI half-width). We will use this sampling proportion to distribute the $B$ observations during stage $r$.

\item The numbers of observations to be taken from populations $X$ and $Y$ in the current stage are $b_{x,r} \equiv \lfloor \gamma_r B \rceil$ and $b_{y,r} \equiv B-b_{x,r}$, respectively, where $\lfloor \cdot \rceil$ is the rounding (nearest integer) operator.

\item Perform the sampling and obtain the results for both samples (including the new number of positive/negative (i.e. 1 or 0) observations for both populations), and then compute the updated $\bar{X}_{r}$ and $\bar{Y}_{r}$. The new total numbers of observations taken just after stage $r$ are now $m_{x,r} = m_{x,r-1}+\lfloor \gamma_r B \rceil$ and $m_{y,r} = m_{y,r-1}+B-\lfloor \gamma_r B \rceil$.

\item Repeat these computations for subsequent decision epochs $r+1, r+2,\ldots$ until the desired half-width bound $\epsilon$ is obtained.

\end{enumerate}

\begin{algorithm}
\caption{Sequential Batching Procedure that minimizes the total sampling cost}\label{Algo:BatchSeqMinCost}
\textbf{Notation:} $m_{x,r},  m_{y,r}$: cumulative numbers of observations taken from populations $X$ and $Y$, respectively, at the completion of stage $r$. 
$\bar{X}_r, \bar{Y}_r$: sample means at the completion of stage $r$.
\begin{algorithmic}[1]
\Require initial sample sizes $m_{x,1}, m_{y,1}$; initial estimates $\hat{p}_x$ and $\hat{p}_y$; desired coverage probability $(1-\alpha)$; sampling capacity (batch size) $B$; desired half-width $\epsilon$; and individual observation costs $c_x>0,  c_y>0$  from populations $X$ and $Y$ 
\State $\tau_{x,1} \gets \hat{p}_x(1-\hat{p}_x)$, $\tau_{y,1} \gets \hat{p}_y(1-\hat{p}_y) $
\State $r \gets 2$, Continue $\gets$ $True$
\While{Continue == $True$}
    \State $\lambda_r \gets \bigg(\dfrac{\sqrt{\tau_{x,r-1} c_x}+\sqrt{\tau_{y,r-1} c_y}}{\epsilon^2/z_{\alpha/2}^2} \bigg)^2$
    \State $\zeta_{x,r} = \sqrt{\dfrac{c_x}{\lambda_r \tau_{x,r-1}}}$,
    $\zeta_{y,r} = \sqrt{\dfrac{c_y}{\lambda_r \tau_{y,r-1}}}$
    \State $\gamma_r = \dfrac{\max(\lceil 1/\zeta_{x,r} \rceil,m_{x,r-1}) - m_{x,r-1}}{(\max(\lceil 1/\zeta_{x,r} \rceil,m_{x,r-1}) - m_{x,r-1})+(\max(\lceil 1/\zeta_{y,r} \rceil,m_{y,r-1}) - m_{y,r-1})} $ 
    \State Sample $\lfloor \gamma_r B \rceil$ observations from population $X$ and $B-\lfloor \gamma_r B \rceil$ from population $Y$
    \State $m_{x,r} \gets m_{x,r-1}+\lfloor \gamma_r B \rceil$, \;
    $m_{y,r} \gets m_{y,r-1}+B-\lfloor \gamma_r B \rceil$
    \State Compute $\bar{X}_r$ and $\bar{Y}_r$
    \State $\tau_{x,r}\gets \bar{X}_r(1-\bar{X}_r) $, \; $\tau_{y,r}\gets \bar{Y}_r(1-\bar{Y}_r) $
    \State $H_r\gets z_{\alpha/2}\sqrt{ \dfrac{\tau_{x,r}}{m_{x,r}} + \dfrac{\tau_{y,r}}{m_{y,r}}}$
    \If{$H_r\leq \epsilon$}
        \State Continue $\gets$ $False$
    \Else
        \State $r \gets r+1$
    \EndIf
\EndWhile
\Ensure Confidence Interval $\bar{X}_r - \bar{Y}_r \pm H_r$
\end{algorithmic}
\end{algorithm}

It is useful to point out that so far we have presented  ``greedy'' algorithms that compute the numbers of observations to be taken from each population in each decision epoch --- by approximately solving the integer optimization problem described by \eqref{LP1.1}--\eqref{LP1.3}, using the cumulative results up to the end of the last epoch as input. Algorithms \ref{Algo:2StageMinCost} and \ref{Algo:BatchSeqMinCost} require  initial estimates $\hat{p}_x$ and $\hat{p}_y$ that are in the open interval $(0,1)$ (else $\tau_{x,r}$ and/or $\tau_{y,r}$ will be zero, which may render the problem trivial). To avoid that issue, both algorithms can instead use the corresponding {\em minimax\/} estimators for $p_x$ and $p_y$, since those estimates are always in $(0,1)$ and avoid the need for an initial sample or a guess on the parameters \citep{Sh21}.

\section{$\ell$-Stage Look-Ahead Algorithm}\label{sec:lstage}

Algorithm \ref{Algo:BatchSeqMinCost} takes into account past and present information in order to carry out its sampling. Since the problem under consideration requires sequential decision-making, it is also interesting to think about algorithms that look
at possible future consequences of the current actions. To this end, we consider a fixed batching capacity of $B$, and we let:

\begin{itemize}
    \item $w_x, w_y$ denote the numbers of successes observed so far from populations $X, \ Y$ respectively, and $m_x, m_y$ represent the numbers of observations taken (so far) from populations $X$ and $Y$.

    \item $b_x, b_y$ denote the possible allocation of observations from populations $X, \ Y$ over the next sampling stage, where $b_x+b_y = B$.
\end{itemize}

We propose next an $\ell$-stage {\em look-ahead\/} procedure that computes the optimal policy $(b_x,b_y)$ for a state $s=(w_x,m_x,w_y,m_y)$ by considering the next $\ell$ decision epochs. The goal of this algorithm is to minimize the expected sampling cost to obtain the desired CI, along with the policy (for the next $\ell$ decision epochs) that achieves this goal. Because we are only looking to the next $\ell$ decision epochs and we may need more observations to obtain the desired CI, this means that we need a function that estimates the ``cost-to-go'' (CTG, i.e., the cost that would need to be incurred to get the desired CI and finish sampling) at a particular sampling stage. In order to obtain such a CTG approximation, we will use Algorithm \ref{Algo:2StageMinCost} to determine an approximate number of observations to be taken from each population to finish the sampling, and then compute the cost associated with those observations.

By using Algorithm \ref{Algo:2StageMinCost} as the basis for our CTG approximation, for each policy (set of sampling budget allocations for the next $\ell$ stages) we can easily compute the expected sampling cost over all possible realizations of the observations given that policy. For each realization we only need to sum the sampling costs associated with the observations taken over the $\ell$ stages (where, in case we obtain the desired CI at any stage, we simply stop sampling); and then at the final stage (after the $\ell$ stages for which our policy gives the allocation), add to those sampling costs the approximate ``cost-to-go'' if the sampling is yet not complete.

For example, when $\ell=1$, we compute the expected cost $C(s)$ starting in state $s$ as follows,
\[
E_{(b_x,b_y)}[C(s)] \; = \; \sum_{s'} P\big(s'|s,b_x,b_y\big) {\rm CTG}(s'),
\]
where the states $s'$ are all the possible sample states resulting after sampling the new $(b_x,b_y)$ observations in the next stage, and $P(s'|s,b_x,b_y)$ is the probability of going to state $s'$ given that we are currently in state $s$ and follow the observation-allocation policy $(b_x,b_y)$. That probability can be computed by using the binomial distribution and any point estimators for $p_x,  p_y$; in particular, we will use in the following sections the sample average success point estimators associated with state $s$; i.e., $\bar{X}=w_x/m_x,\bar{Y}=w_y/m_y$. The expectation is computed for all $(b_x,b_y)$ such that $b_x+b_y=B$, and ${\rm CTG}(s')$ is the cost-to-go of state $s'$ given by the optimal solution obtained from Algorithm \ref{Algo:2StageMinCost} applied to state $s'$.

Similarly, when $\ell=2$, we have
\begin{eqnarray*}
E_{(b_x,b_y),(b_x^\prime,b_y^\prime)}[C(s)]
& = & \sum_{s''} \sum_{s'} P\big(s'|s,b_x,b_y\big) P\big(s''|s',b_x^\prime,b_y^\prime\big) {\rm CTG}(s'') \\
& = & \sum_{s'} P\big(s'|s,b_x,b_y\big) E_{(b_x^\prime,b_y^\prime)}[C(s')],
\end{eqnarray*}
where $b_x+b_y=b_x^\prime+b_y^\prime=B$, and $(b_x,b_y)$ represents the numbers of observations to be taken from populations $X,\ Y$ respectively, during the next stage, and $(b_x',b_y')$ similarly for the subsequent stage. The procedure for $\ell = 1$ is given in Algorithm \ref{Algo:1StageLookAhead}.

\begin{algorithm}
\caption{One-Step Look-Ahead Procedure to minimize total sampling costs}\label{Algo:1StageLookAhead}
\textbf{Notation:} $m_{x,r}, m_{y,r}$: cumulative numbers of observations taken from populations $X$ and $Y$, respectively, at the completion of stage $r$. $\bar{X}_r, \bar{Y}_r$: sample means at the completion of stage $r$.
\begin{algorithmic}[1]
\Require initial sample sizes $m_{x,1},  m_{y,1}$; initial estimates $\hat{p}_x$ and $\hat{p}_y$; desired coverage probability $(1-\alpha)$; sampling capacity $B$; desired half-width $\epsilon$; and individual observation costs $c_x>0, \ c_y>0$  from populations $X$ and $Y$.
\State $\tau_{x,1} \gets \hat{p}_x(1-\hat{p}_x), \, \tau_{y,1} \gets \hat{p}_y(1-\hat{p}_y) $
\State $r \gets 2$, Continue $\gets$ $True$
\While{Continue == $True$}
    \For {all policies $(b_x,b_y)$ with $b_x+b_y=B$}
        \State $E_{(b_x,b_y)}\gets 0$
        \For{all pairs $(i,j)$ with $0\leq i \leq b_x$ and $0\leq j\leq b_y$}
            \State Execute Algorithm \ref{Algo:2StageMinCost} with inputs: $m_{x,r-1}$, $m_{y,r-1}$, $\bar{X}_{r-1}$, $\bar{Y}_{r-1}$, $(1-\alpha)$, $\epsilon$, $c_x$, $c_y$
            \State Compute from step 7's output the sampling costs, denoted CTG, of the second stage,
            \State $E_{(b_x,b_y)}\gets E_{(b_x,b_y)} + \mbox{CTG} \cdot  {b_{x} \choose i} \bar{X}_{r-1}^i (1-\bar{X}_{r-1})^{b_{x}-i}  {b_{y} \choose j} \bar{Y}_{r-1}^j (1-\bar{Y}_{r-1})^{b_{y}-j} $ 
        \EndFor
    \EndFor
    \State $(b_x^\star,b_y^\star) \gets \text{argmin}_{(b_x,b_y)}E_{(b_x,b_y)}$
    \State Sample $b_x^\star, b_y^\star$ observations from populations $X $ and $ Y$, respectively 
    \State $m_{x,r} \gets m_{x,r-1}+b_x^\star, \, m_{y,r} \gets m_{y,r-1}+b_y^\star$
    \State Compute $\bar{X}_r$ and $\bar{Y}_r$
    \State $\tau_{x,r}\gets \bar{X}_r(1-\bar{X}_r) , \, \tau_{y,r}\gets \bar{Y}_r(1-\bar{Y}_r)$
    \State $H_r\gets z_{\alpha/2}\sqrt{ \dfrac{\tau_{x,r}}{m_{x,r}} + \dfrac{\tau_{y,r}}{m_{y,r}}}$
    \If{$H_r \leq \epsilon$}
        \State Continue $\gets$ $False$
    \Else
        \State $r \gets r+1$
    \EndIf
\EndWhile
\Ensure Confidence Interval $\bar{X}_r - \bar{Y}_r \pm H_r$
\end{algorithmic}
\end{algorithm}

Clearly, when the number of look-ahead stages $\ell$ increases, we will obtain an expected value that becomes closer to the true expected cost at stage $s$ (we are only approximating because of the use of the CTG), as there are fewer realizations where we need to add the CTG approximation (which follows from the fact that by considering more stages, more observations are taken, and the proportion of times the sampling ends by the $\ell$ stages increases). Since the number of policies is $O(B)$ at each possible decision point, we see that the total number of expectations to compute is $O(B^{\ell})$, which makes it increasingly difficult to find the optimal policy as $\ell$ increases. In fact, when $\ell$ goes to infinity, this algorithm becomes equivalent to an intractable MDP formulation that minimizes expected costs, presented in Appendix \ref{sec:MDP}.

\section{Experimental Results}
\label{sec:results}

This section presents the results of our Monte Carlo experimentation. In particular, the goal of \S\ref{subsec:results:comparing-procedures} is to illustrate the value of the procedures that minimize sampling costs. In order to do so, the procedures proposed in this paper (Algorithms \ref{Algo:2StageMinCost} and \ref{Algo:BatchSeqMinCost}) are benchmarked against their counterpart procedures that minimize the total number of observations taken (see Algorithms \ref{Algo:2StageMinSamples}--\ref{Algo:BatchSeqMinSamples}  in Appendix \ref{App}, reproduced from \cite{erazo21}) and the baseline procedures depicted in Algorithms \ref{Algo:Conservative} and \ref{Algo:NaiveSeqMinCost}. 
Table \ref{table:algorithms} summarizes the purpose of each of the various algorithms discussed in the paper.

\begin{table}[h]
		\centering
		\begin{tabular}{|c|c|c|}
			\hline
		Algorithm  &  Description \\ \hline
		\ref{Algo:Conservative} & Conservative two-stage procedure\\
		\ref{Algo:2StageMinCost} & Two-stage procedure to minimize cost \\
		\ref{Algo:NaiveSeqMinCost} & Naive sequential procedure to minimize cost \\ 			\ref{Algo:BatchSeqMinCost} & Batched sequential procedure to minimize cost \\
		\ref{Algo:1StageLookAhead} & One-step look-ahead procedure to minimize cost\\ \hline
		 \ref{Algo:2StageMinSamples}  & Two-stage procedure to minimize number of observations  \\			
		\ref{Algo:FullySeqMinSamples} & Fully sequential procedure to minimize number of observations \\
		\ref{Algo:BatchSeqMinSamples} & Batched sequential procedure to minimize number of observations  \\ \hline
		\end{tabular}
		\caption{List of algorithms (Algorithms  \ref{Algo:2StageMinSamples}--\ref{Algo:BatchSeqMinSamples} are given in the Appendix).}
		\label{table:algorithms}
	\end{table}

The comparisons are carried out under the three following sampling schemes (within a particular scheme below, the algorithms are arranged in the order: benchmark algorithm vs.\ min.\ observations algorithm vs.\ min.\ cost algorithm): 
\begin{itemize}
\item two-stage (Algorithm \ref{Algo:Conservative} vs.\ Algorithm \ref{Algo:2StageMinSamples} vs Algorithm \ref{Algo:2StageMinCost});
\item fully sequential (Algorithm 
\ref{Algo:NaiveSeqMinCost} vs.\ Algorithm \ref{Algo:FullySeqMinSamples} vs.\ Algorithm \ref{Algo:BatchSeqMinCost}); and
\item batched sampling with $B=10$ (Algorithm \ref{Algo:NaiveSeqMinCost} vs.\ Algorithm \ref{Algo:BatchSeqMinSamples} vs.\ Algorithm \ref{Algo:BatchSeqMinCost}).
\end{itemize}

\S\ref{subsec:results:comparing-cost-minimization-schemes} discusses the differences between the proposed cost-saving procedures under the different sampling schemes (two-stage, fully sequential, and batched) and the one-step look-ahead procedure defined in \S\ref{sec:lstage}. Finally, \S \ref{sec:studycase} presents a detailed case study in which we validate the benefits that can be obtained by using Algorithm \ref{Algo:BatchSeqMinCost} on a healthcare application.

\subsection{Cost-Savings Procedures vs.\ Other Procedures under Different Sampling Schemes} 
\label{subsec:results:comparing-procedures}


Tables \ref{table:ResultsOneStageComparison}, \ref{table:ResultsFullySequentialComparison}, and \ref{table:ResultsBatchingComparison} compare the cost-saving procedures versus the procedures from \cite{erazo21} that are presented in Appendix \ref{App} and the baseline procedures from Algorithms \ref{Algo:Conservative} and \ref{Algo:NaiveSeqMinCost}, for each of the sampling schemes (two-stage, fully sequential, and batched). 
We run nine different experiments/scenarios, where each experiment has $R=1000$ replications each incorporating initial sample sizes of 50 (used to compute $\hat{p}_x$ and $\hat{p}_y$)  for both populations $X$ and $Y$. Each experiment is determined by a vector with four values, $({\cal E},c_x/c_y,p_x,p_y)$, given in column 1 of each table, where ${\cal E}$ is the generic designation of the experiments/scenarios that will be used over the discussion. Note that, without loss of generality, we can simply work with the cost ratio $c_x/c_y$ instead of the two individual costs. All experiments use $\epsilon=0.05$ and $\alpha=0.05$.

In column 2 of these tables, the label ``Coverage Probability (\%)'' denotes the empirical percentage $\hat{c}$ of CIs that actually covered $p_x-p_y$; this entry has generic standard error $\sqrt{\hat{c}(1-\hat{c})/R}$.  The label ``Half-width Achieved (\%)'' denotes the empirical proportion of replications that actually attained the half-width requirement.  Note that Algorithm \ref{Algo:Conservative} is a conservative two-stage procedure that {\em always\/} attains the desired width, while Algorithms \ref{Algo:2StageMinCost} and \ref{Algo:2StageMinSamples} are two-stage procedures that {\em do not guarantee\/} that requirement. All of the other algorithms discussed in these tables are designed to guarantee the half-width requirement. The term ``Cost Gap \%'' is
defined/normalized for a generic procedure as follows.  First of all, for replication $i=1,2,\ldots,R$, define the cost gap for a procedure with respect to the corresponding naive (benchmark) procedure by $\mbox{Gap}_i(\mbox{Procedure}) = \mbox{Cost}_i(\mbox{Procedure}) / \mbox{Cost}_i(\mbox{Naive})$.  Because Gap$_i$ is a ratio, the overall gap measure  reported in the table is the geometric average, i.e.,  $\mbox{Cost Gap \%} = (\prod_{i=1}^R {\rm Gap}_{i})^{1/R}$. Gap calculations are based only on the observations taken outside of the initial estimation stage. We also report the standard deviation of the Gap$_i$'s taken over all of the $R$ replications, denoted ``std.''.  Column 2 also lists the maximum and minimum gaps from the $R$ replications as performance measures.  The performance results for three related algorithms are presented in columns 3--5 of each table.  Table \ref{table:ResultsOneStageComparison} has results for three two-stage procedures; Table \ref{table:ResultsFullySequentialComparison} for three
fully sequential procedures; and Table  \ref{table:ResultsBatchingComparison} for three procedures employing batched sampling with $B=10$.  Among the three procedures appearing in any table, column 3 represents the baseline algorithm; column 4 the algorithm designed to minimize the number of observations; and column 5 the algorithm designed to minimize costs.

We point out some details about the experiments.  Table \ref{table:list-of-scenarios} lists the scenarios on which our Monte Carlo experiments are based.  
	\begin{table}[!h]
		\centering
		\begin{tabular}{|c|c|}
			\hline
		$({\cal E}, c_x/c_y,p_x,p_y)$  &  comments \\ \hline
		(s1, 1, 0.3, 0.2) & \\
		(s2, 1, 0.5, 0.2) & cost-neutral group of scenarios \\
		(s3, 1, 0.5, 0.5) & \\ 			\hline
		(s4, $\frac{1}{3}$, 0.3, 0.2) & \\
		(s5, $\frac{1}{3}$, 0.5, 0.2) & low cost associated with higher $p_x$\\
		(s6, $\frac{1}{3}$, 0.5, 0.5) & \\ 			\hline
		(s7, 5, 0.3, 0.2) & \\
		(s8, 5, 0.5, 0.2) & high cost associated with higher $p_x$\\
		(s9, 5, 0.5, 0.5) & \\ \hline
		\end{tabular}
		\caption{List of scenarios }
		\label{table:list-of-scenarios}
	\end{table}

  \begin{itemize}
     \item The scenarios can be divided into the following three groups: (s1,s2,s3), (s4,s5,s6), and (s7,s8,s9). The three scenarios within a particular group share the {\em same cost ratio $c_x/c_y$\/}  but have different Bernoulli parameters, namely, $(p_x,p_y) = (0.3,0.2)$, $(0.5,0.2)$, and $(0.5,0.5)$. Note that because of the Bernoulli distribution's symmetry (and the definitions of $\tau_x$ and $\tau_y$), it is not necessary to study pairs with $p_x$ or $p_y$ greater than 0.5.
     \item The scenarios can also be divided into a different collection of three groups: (s1,s4,s7), (s2,s5,s8), and (s3,s6,s9). The three scenarios comprising each of these groups share the same $(p_x,p_y)$-values for the Bernoulli parameters, but sampling costs differ for each of those three scenarios. The first scenario in each group has a cost ratio of 1 (i.e., equal costs, $c_x=c_y$), while the other two scenarios within a group have cost ratios of $\frac{1}{3}$ and $5$, respectively.
 \end{itemize}
 
 We dive into the results along with their interpretations:
 
 \begin{itemize}
     \item When $c_x=c_y$ (scenarios s1, s2, s3), it is obvious that minimizing the number of observations taken is equivalent to minimizing the total cost of the sampling scheme (cf.\ Equation \eqref{LP1.1}, in which case we minimize $m_{x,2}+m_{y,2}$). On the other hand, if $p_x=p_y$ (scenarios s3, s6, s9), then sampling from both populations at the same rate actually minimizes the total number of observations taken, which follows from the symmetry of Equation \eqref{LP1.2} and the fact that minimizing the number of observations taken is equivalent to considering an objective function of $m_{x,2}+m_{y,2}$; thus, for cases (s3, s6, s9) the naive sequential method from Algorithm \ref{Algo:NaiveSeqMinCost} and the sequential algorithms from Appendix \ref{App} are equivalent. When both conditions $c_x=c_y$ and $p_x=p_y$  are true (scenario s3), then in each of Tables \ref{table:ResultsOneStageComparison}, \ref{table:ResultsFullySequentialComparison}, and \ref{table:ResultsBatchingComparison}, the three algorithms that are compared in that respective table are all equivalent. These remarks are corroborated experimentally in Tables  \ref{table:ResultsOneStageComparison}, \ref{table:ResultsFullySequentialComparison}, and \ref{table:ResultsBatchingComparison}. 
     
     With respect to the results for scenarios s1 and s2 under the two-stage sampling scheme, the cost is reduced by over 20\%. Furthermore, for scenarios s1 and s2 under the sequential sampling schemes, the methods that minimize the number of observations taken (fourth column) and the methods that minimize the total cost (fifth column) have significantly smaller costs than the ``naive'' sequential method (Algorithm \ref{Algo:NaiveSeqMinCost}), even though the differences in the gap between these two methods and the ``naive'' method are rather minor ($\leq 2\%$).
      
     \item Focusing on Table \ref{table:ResultsOneStageComparison}, we note that the Coverage Probability and Half-width Achieved \% are equal for  Algorithms \ref{Algo:Conservative} and  \ref{Algo:2StageMinSamples}, when we look at the same value for the Bernoulli parameters (i.e., looking at groups (s1,s4,s7), (s2,s5,s8), and (s3,s6,s9)); but this does not hold true for the procedure that minimizes costs (Algorithm \ref{Algo:2StageMinCost}). This makes sense as Algorithm \ref{Algo:Conservative} and Algorithm \ref{Algo:2StageMinSamples} do not take into account the cost ratio for decision-making, whereas Algorithm \ref{Algo:2StageMinCost} does --- so that its performance is not completely determined by the Bernoulli parameters alone.  
     
     The previous equality results for the coverage probability and half-width are not repeated in Tables \ref{table:ResultsFullySequentialComparison} and \ref{table:ResultsBatchingComparison}. This stems from the fact that the fully sequential and batching schemes are affected by the inherent randomness of the observations and their ordering (they rely on the sample paths obtained), whereas the two-stage sampling scheme does not have this issue.

     \item It is easy to see from Tables \ref{table:ResultsFullySequentialComparison} and \ref{table:ResultsBatchingComparison} that in scenario groups (s1,s2,s3) and (s4,s5,s6), the cost gap for the fourth column (procedures minimizing the number of observations taken) and the fifth column (procedures minimizing total costs) increases when $p_x$ and $p_y$ are farther apart (i.e., the gap of s6 $>$ gap s4 $>$ gap s5), whereas the opposite is true for the group (s7,s8,s9) (where gap s9 $<$ gap s7 $<$ gap s8). This suggests that when the population with the higher sampling cost has a smaller [larger] success probability than the other population, then more discrepancy increases [reduces] the gap. Such behavior with respect to geometric gaps does not occur in Table \ref{table:ResultsOneStageComparison} because the benchmark is versus the conservative approach that does not vary on costs; and so the gap is only reduced when $p_x$ and $p_y$ get closer to 0.5 (and the number of observations taken approaches what happens in the conservative approach).

     \item With respect to the geometric gaps, we notice that in all the tables of this section, the procedures minimizing costs (column 5) induce a significantly smaller gap compared to the baseline methods (column 3) for all scenarios, except for the one scenario where they are equivalent (s3, under sequential sampling schemes); and a significantly smaller gap than the procedure that minimizes the number of observations taken for all scenarios except for the ones where costs are equivalent (s1,s2,s3). In general, when looking at Tables \ref{table:ResultsFullySequentialComparison} and \ref{table:ResultsBatchingComparison} the gaps are very similar within equal scenarios (e.g., comparing scenario s5 results in both tables), and an improvement of up to around 15\% can be obtained vs.\ the naive benchmark procedure (Algorithm \ref{Algo:NaiveSeqMinCost}). With respect to the procedures that minimize observations (column 4), when the population with the higher cost has a smaller [higher] probability parameter, then the cost is significantly smaller [somewhat higher] than that of the baseline; and this means that under those circumstances reducing the number of observations has as a trade-off an increase in the overall cost.

     \item Furthermore, with respect to maximum and minimum gaps, under the two-stage sampling scheme (Table \ref{table:ResultsOneStageComparison}), the two-stage procedures to minimize observations taken and to minimize costs (columns 4 and 5, respectively) never have a cost that exceeds that of the conservative approach (column 3); however, those savings also result due to the already discussed inherent problematic issue underlying two-stage procedures: most of the time the desired half-width is not obtained, as can be seen in the ``Half-width Achieved \%'' rows.
    
     \item When looking solely at the two-stage sampling scheme (Table \ref{table:ResultsOneStageComparison}), the overall Coverage Probability and Half-width Achieved \% (over all 9 scenarios) of the conservative approach (Algorithm \ref{Algo:Conservative}), the two-stage procedure that minimizes observations (Algorithm \ref{Algo:2StageMinSamples}), and the two-stage procedure minimizing costs (Algorithm \ref{Algo:2StageMinCost}) are (94.0, 95.6, 94.8) (compared to the desired 95\%) and (100.0, 32.7, 33.0), respectively. This shows that our proposed algorithms can produce a significant increase (standard errors from the 1000 replications are below 0.01) in the coverage probability (which is also backed up by the fact that the proposed two-stage algorithms beat the conservative approach in all of the scenarios), while substantially reducing the sampling costs (column 5 has up to 50\% of cost reductions, and a minimum of 2\%), at the expense of a decrease in the probability of obtaining the desired half-width.   
     
     \item When looking at Tables \ref{table:ResultsFullySequentialComparison} and \ref{table:ResultsBatchingComparison}, columns 3, 4, and 5,we obtain the following ``overall'' coverage probabilities by averaging over the nine scenarios:  (94.81, 94.7, 95.1) and (95.2, 95.0, 94.6), which tend to be near the nominal level while also satisfying the desired half-width. This suggests that using the cost-saving proposed algorithms (column 5) {significantly reduces} the sampling costs while preserving similar coverage probabilities and satisfying the half-width criterion.
 \end{itemize}

In terms of an overall conclusion from these comparisons, it is clear that if the goal is to minimize the sampling costs, the cost-saving procedures proposed over the course of this paper are the most-appropriate choices for their respective sampling-scheme (two-stage, fully sequential, or batches). Further, for the sequential sampling schemes, these procedures induce a similar probability of obtaining the desired CI half-width as well as coverage probabilities near the nominal level.

	\begin{table}[!h]
		\centering
		\begin{tabular}{|c|c|c|c|c|}
			\hline
				$({\cal E}, c_x/c_y,p_x,p_y)$ &  Performance Measures & Algorithm \ref{Algo:Conservative} & Algorithm \ref{Algo:2StageMinSamples} & Algorithm \ref{Algo:2StageMinCost}    \\
					 &   & conservative & min.\ obs.\ & min.\ cost   \\
			\hline \hline
			\multirow{4}{*}{(s1, 1, 0.3, 0.2)} & Coverage Probability (\%)	  & 93.9 & 94.7 & 93.9 \\
			& Half-width Achieved (\%) & 100 & 43.3 & 44.5  \\
			& Cost Gap \% (std.) & 100 (--) & 69.1 (9.3) & 69.1 (9.3) \\
			& Max Gap \% (Min) & --- & 92.8 (30.2) & 92.8 (30.2) \\
			\hline
			\multirow{4}{*}{(s2, 1, 0.5, 0.2)} & Coverage Probability (\%)	  & 94.1 & 96.9 & 95.6 \\
			& Half-width Achieved (\%) & 100 & 44.3 & 45.5  \\
			& Cost Gap \% (std.) & 100 (--) & 77.2 (8.1) & 77.2 (8.1) \\
			& Max Gap \% (Min) & --- & 97.1 (36.9) & 97.1 (36.9) \\
			\hline
			\multirow{4}{*}{(s3, 1, 0.5, 0.5)} & Coverage Probability (\%)	  & 93.9 & 95.2 & 94.6 \\
			& Half-width Achieved (\%) & 100 & 10.5 & 10.2  \\
			& Cost Gap \% (std.) & 100 (--) & 97.9 (2.0) & 97.9 (2.0) \\
			& Max Gap \% (Min) & --- & 100.0 (83.0) & 100.0 (83.0) \\
			\hline
			\hline
			\multirow{4}{*}{(s4, $\frac{1}{3}$, 0.3, 0.2)} & Coverage Probability (\%)	  & 93.9 & 94.7 & 94.5 \\
			& Half-width Achieved (\%) & 100 & 43.3 & 44.8  \\
			& Cost Gap \% (std.) & 100 (--) & 66.3 (10.1) & 61.3 (9.5) \\
			& Max Gap \% (Min) & --- & 93.0 (20.4) & 86.3 (18.6) \\
			\hline
			\multirow{4}{*}{(s5, $\frac{1}{3}$, 0.5, 0.2)} & Coverage Probability (\%)	  & 94.1 & 96.9 & 95.2 \\
			& Half-width Achieved (\%) & 100 & 44.3 & 48.0  \\
			& Cost Gap \% (std.) & 100 (--) & 72.3 (9.8) & 66.7 (9.2) \\
			& Max Gap \% (Min) & --- & 97.4 (24.5) & 90.5 (22.6) \\
			\hline
			\multirow{4}{*}{(s6, $\frac{1}{3}$, 0.5, 0.5)} & Coverage Probability (\%)	  & 93.9 & 95.2 & 94.5 \\
			& Half-width Achieved (\%) & 100 & 10.5 & 9.3  \\
			& Cost Gap \% (std.) & 100 (--) & 97.9 (2.1) & 90.9 (2.0) \\
			& Max Gap \% (Min) & --- & 100.0 (79.9) & 92.8 (73.9) \\
			\hline
			\hline
			\multirow{4}{*}{(s7, 5, 0.3, 0.2)} & Coverage Probability (\%)	  & 93.9 & 94.7 & 94.4 \\
			& Half-width Achieved (\%) & 100 & 43.3 & 44.3  \\
			& Cost Gap \% (std.) & 100 (--) & 72.6 (8.9) & 63.0 (7.7) \\
			& Max Gap \% (Min) & --- & 93.5 (42.2) & 80.9 (35.7) \\
			\hline
			\multirow{4}{*}{(s8, 5, 0.5, 0.2)} & Coverage Probability (\%)	  & 94.1 & 96.9 & 95.2 \\
			& Half-width Achieved (\%) & 100 & 44.3 & 39.5  \\
			& Cost Gap \% (std.) & 100 (--) & 83.8 (6.0) & 49.4 (4.8) \\
			& Max Gap \% (Min) & --- & 96.8 (53.3) & 83.8 (49.4) \\
			\hline
			\multirow{4}{*}{(s9, 5, 0.5, 0.5)} & Coverage Probability (\%)	  & 93.9 & 95.2 & 94.7 \\
			& Half-width Achieved (\%) & 100 & 10.5 & 10.8  \\
			& Cost Gap \% (std.) & 100 (--) & 97.8 (2.1) & 84.5 (1.9) \\
			& Max Gap \% (Min) & --- & 100.0 (85.5) & 86.3 (73.2) \\
			\hline
			\end{tabular}
		\caption{Results for two-stage sampling scheme}
		\label{table:ResultsOneStageComparison}
	\end{table}

\begin{table}[!h]
		\centering
		\begin{tabular}{|c|c|c|c|c|}
			\hline
				$({\cal E}, c_x/c_y,p_x,p_y)$  &  Performance Measures & Algorithm \ref{Algo:NaiveSeqMinCost} & Algorithm \ref{Algo:FullySeqMinSamples} & Algorithm \ref{Algo:BatchSeqMinCost}    \\
					 &  & naive seq.\ & min.\ obs.\ &  min.\ cost   \\
			\hline \hline
			\multirow{4}{*}{(s1, 1, 0.3, 0.2)} & Coverage Probability (\%)	  & 94.8 & 93.7 & 95.1 \\
			& Half-width Achieved (\%) & 100 & 100 & 100  \\
			& Cost Gap \% (std.) & 100 (--) & 99.1 (5.4) & 99.2 (5.4) \\
			& Max Gap \% (Min) & --- & 118.8 (80.5) & 113.8 (83.2) \\
			\hline
			\multirow{4}{*}{(s2, 1, 0.5, 0.2)} & Coverage Probability (\%)	  & 95.2 & 95.8 & 96.1 \\
			& Half-width Achieved (\%) & 100 & 100 & 100  \\
			& Cost Gap \% (std.) & 100 (--) & 98.3 (3.8) & 98.3 (3.9) \\
			& Max Gap \% (Min) & --- & 110.2 (85.5) & 111.9 (85.4) \\
			\hline
			\multirow{4}{*}{(s3, 1, 0.5, 0.5)} & Coverage Probability (\%)	  & 94.5 & 94.0 & 95.4 \\
			& Half-width Achieved (\%) & 100 & 100 & 100  \\
			& Cost Gap \% (std.) & 100 (--) & 100.0 (0.2) & 100.0 (0.2) \\
			& Max Gap \% (Min) & --- & 101.0 (98.9) & 100.8 (99.2) \\
			\hline
			\hline
			\multirow{4}{*}{(s4, $\frac{1}{3}$, 0.3, 0.2)} & Coverage Probability (\%)	  & 94.8 & 94.4 & 95.1 \\
			& Half-width Achieved (\%) & 100 & 100 & 100  \\
			& Cost Gap \% (std.) & 100 (--) & 95.3 (5.7) & 88.0 (5.6) \\
			& Max Gap \% (Min) & --- & 114.8 (74.8) & 105.8 (69.6) \\
			\hline
			\multirow{4}{*}{(s5, $\frac{1}{3}$, 0.5, 0.2)} & Coverage Probability (\%)	  & 95.2 & 95.6 & 95.0 \\
			& Half-width Achieved (\%) & 100 & 100 & 100  \\
			& Cost Gap \% (std.) & 100 (--) & 92.3 (4.3) & 85.1 (4.5) \\
			& Max Gap \% (Min) & --- & 105.1 (78.0) & 98.4 (70.8) \\
			\hline
			\multirow{4}{*}{(s6, $\frac{1}{3}$, 0.5, 0.5)} & Coverage Probability (\%)	  & 94.5 & 94.3 & 94.2 \\
			& Half-width Achieved (\%) & 100 & 100 & 100  \\
			& Cost Gap \% (std.) & 100 (--) & 100.0 (0.2) & 92.8 (0.2) \\
			& Max Gap \% (Min) & --- & 101.1 (98.7) & 93.6 (91.7) \\
			\hline
			\hline
			\multirow{4}{*}{(s7, 5, 0.3, 0.2)} & Coverage Probability (\%)	  & 94.9 & 94.0 & 94.6 \\
			& Half-width Achieved (\%) & 100 & 100 & 100  \\
			& Cost Gap \% (std.) & 100 (--) & 104.1 (5.3) & 90.5 (4.8) \\
			& Max Gap \% (Min) & --- & 124.6 (88.0) & 109.1 (75.8) \\
			\hline
			\multirow{4}{*}{(s8, 5, 0.5, 0.2)} & Coverage Probability (\%)	  & 95.4 & 95.5 & 95.1 \\
			& Half-width Achieved (\%) & 100 & 100 & 100  \\
			& Cost Gap \% (std.) & 100 (--) & 106.4 (3.3) & 93.0 (2.7) \\
			& Max Gap \% (Min) & --- & 118.1 (95.5) & 102.8 (85.7) \\
			\hline
			\multirow{4}{*}{(s9, 5, 0.5, 0.5)} & Coverage Probability (\%)	  & 94.3 & 94.6 & 95.4 \\
			& Half-width Achieved (\%) & 100 & 100 & 100  \\
			& Cost Gap \% (std.) & 100 (--) & 100.0 (0.2) & 86.4 (0.2) \\
			& Max Gap \% (Min) & --- & 101.1 (98.8) & 87.0 (84.9) \\
			\hline
			\end{tabular}
		\caption{Results for fully sequential sampling scheme}
		\label{table:ResultsFullySequentialComparison}
	\end{table}

\begin{table}[!h]
		\centering
		\begin{tabular}{|c|c|c|c|c|}
			\hline
				$({\cal E}, c_x/c_y,p_x,p_y)$  &  Performance Measures   & Algorithm \ref{Algo:NaiveSeqMinCost} & Algorithm \ref{Algo:BatchSeqMinSamples} & Algorithm \ref{Algo:BatchSeqMinCost}    \\
					 &  & naive seq.\ & batch min.\ obs.\ &  batch min.\ cost   \\
			\hline \hline
			\multirow{4}{*}{(s1, 1, 0.3, 0.2)} & Coverage Probability (\%)	  & 96.3 & 95.4 & 95.1 \\
			& Half-width Achieved (\%) & 100 & 100 & 100  \\
			& Cost Gap \% (std.) & 100 (--) & 99.3 (5.2) & 99.4 (5.2) \\
			& Max Gap \% (Min) & --- & 114.6 (80.2) & 114.9 (82.4) \\
			\hline
			\multirow{4}{*}{(s2, 1, 0.5, 0.2)} & Coverage Probability (\%)	  & 93.0 & 94.6 & 95.4 \\
			& Half-width Achieved (\%) & 100 & 100 & 100  \\
			& Cost Gap \% (std.) & 100 (--) & 98.7 (3.8) & 98.6 (3.8) \\
			& Max Gap \% (Min) & --- & 109.9 (85.8) & 111.0 (86.3) \\
			\hline
			\multirow{4}{*}{(s3, 1, 0.5, 0.5)} & Coverage Probability (\%)	  & 96.2 & 94.8 & 94.9 \\
			& Half-width Achieved (\%) & 100 & 100 & 100  \\
			& Cost Gap \% (std.) & 100 (--) & 100.0 (0.2) & 100.0 (0.2) \\
			& Max Gap \% (Min) & --- & 100.7 (98.6) & 100.7 (99.3) \\
			\hline
			\hline
			\multirow{4}{*}{(s4, $\frac{1}{3}$, 0.3, 0.2)} & Coverage Probability (\%)	  & 94.7 & 94.6 & 94.5 \\
			& Half-width Achieved (\%) & 100 & 100 & 100  \\
			& Cost Gap \% (std.) & 100 (--) & 95.7 (5.7) & 88.2 (5.6) \\
			& Max Gap \% (Min) & --- & 116.1 (79.5) & 112.0 (71.1) \\
			\hline
			\multirow{4}{*}{(s5, $\frac{1}{3}$, 0.5, 0.2)} & Coverage Probability (\%)	  & 94.9 & 95.9 & 93.7 \\
			& Half-width Achieved (\%) & 100 & 100 & 100  \\
			& Cost Gap \% (std.) & 100 (--) & 92.8 (4.2) & 85.5 (4.5) \\
			& Max Gap \% (Min) & --- & 104.9 (78.7) & 101.8 (70.9) \\
			\hline
			\multirow{4}{*}{(s6, $\frac{1}{3}$, 0.5, 0.5)} & Coverage Probability (\%)	  & 95.9 & 95.0 & 94.3 \\
			& Half-width Achieved (\%) & 100 & 100 & 100  \\
			& Cost Gap \% (std.) & 100 (--) & 100.0 (0.2) & 92.8 (0.2) \\
			& Max Gap \% (Min) & --- & 100.8 (99.1) & 93.6 (91.3) \\
			\hline
			\hline
			\multirow{4}{*}{(s7, 5, 0.3, 0.2)} & Coverage Probability (\%)	  & 94.9 & 94.3 & 95.3 \\
			& Half-width Achieved (\%) & 100 & 100 & 100  \\
			& Cost Gap \% (std.) & 100 (--) & 104.2 (4.9) & 90.5 (4.7) \\
			& Max Gap \% (Min) & --- & 118.6 (89.2) & 106.2 (77.7) \\
			\hline
			\multirow{4}{*}{(s8, 5, 0.5, 0.2)} & Coverage Probability (\%)	  & 95.4 & 95.3 & 94.5 \\
			& Half-width Achieved (\%) & 100 & 100 & 100  \\
			& Cost Gap \% (std.) & 100 (--) & 106.7 (3.4) & 93.1 (2.6) \\
			& Max Gap \% (Min) & --- & 117.6 (94.9) & 103.1 (85.1) \\
			\hline
			\multirow{4}{*}{(s9, 5, 0.5, 0.5)} & Coverage Probability (\%)	  & 95.3 & 94.8 & 93.8 \\
			& Half-width Achieved (\%) & 100 & 100 & 100  \\
			& Cost Gap \% (std.) & 100 (--) & 100.0 (0.2) & 86.2 (0.2) \\
			& Max Gap \% (Min) & --- & 101.4 (99.0) & 87.4 (84.8) \\
			\hline
		    \end{tabular}
		\caption{Results for batch sampling scheme ($B=10$)}
		\label{table:ResultsBatchingComparison}
	\end{table}

\subsection{Comparison of Procedures that Minimize Cost}
\label{subsec:results:comparing-cost-minimization-schemes}

Now we move on to evaluate the performance of the different cost-minimization procedures and corresponding sampling schemes (Algorithms \ref{Algo:2StageMinCost}, \ref{Algo:BatchSeqMinCost}, and \ref{Algo:1StageLookAhead}). Table \ref{table:ResultsCostComparison1} gives results to enable a comparison for the same nine scenarios previously described, each for 1000 replications, for the two-stage (Algorithm \ref{Algo:2StageMinCost}), fully sequential (Algorithm \ref{Algo:BatchSeqMinCost} with $B=10$), batched (Algorithm \ref{Algo:BatchSeqMinCost} with $B=10$), and one-step look-ahead (Algorithm \ref{Algo:1StageLookAhead} with $B=10$) procedures. It is easy to see that under the initial samples sizes ($m_{x,1}=m_{y,1}=50$) the two-stage procedure has the smallest cost, and the gap between the sequential methods and the two-stage method is usually between $2\%$ and 3\%. Of course, as already mentioned, the two-stage method does not guarantee the desired half-width (in fact, there is no scenario in our study in which the desired half-width is obtained even 50\% of the time). The two-stage procedure also has the lowest coverage probability in six of the nine scenarios (including all three with large discrepancies between $p_x$ and $p_y$) and the lowest overall coverage probability (of 93.8 over the nine scenarios, which is well below the specification). It is also easy to see that the fully sequential algorithm outperforms the other two in all scenarios on average by a small fraction, but at the expense on requiring approximately 10 times as many decision epochs. Both the batching method and the one-step-ahead sampling scheme have the same cost and need approximately the same number of decision epochs.  However, the batching method has a more-consistent coverage probability (overall 95.1, slightly greater than nominal) and is considerably easier to compute (at least 500 times faster); and as such should be preferred over the one-step-ahead sampling scheme (the performance of a  higher-order-step-ahead procedure could be studied in the future). As a recap, moderate batching almost always results in coverages, half-widths, and costs that are better than or at least more-or-less the equal to any of the methods --- but is faster in terms of the number of stages and/or computation time required compared to other competitive methods (e.g., fully sequential or one-step look-ahead).

\begin{table}[!h]
{\footnotesize
		\centering
		\begin{tabular}{|c|c|c|c|c|c|}
			\hline
				$({\cal E}, c_x/c_y,p_x,p_y)$  & Performance Measures & Algorithm \ref{Algo:2StageMinCost} & Algorithm \ref{Algo:BatchSeqMinCost}
				& Algorithm \ref{Algo:BatchSeqMinCost}
				& Algorithm \ref{Algo:1StageLookAhead} \\
					 &     & Two-stage & Fully seq. & Batches & One-step ahead  \\
			\hline \hline
			\multirow{5}{*}{(s1, 1, 0.3, 0.2)} & Coverage Probability (\%)	  & 93.6 & 94.5 & 95.5 & 94.2 \\
			& Half-width Achieved (\%) & 43.0 & 100 & 100 & 100 \\
			& Cost Gap \% (std.) & 100 (--) & 102.7 (15.0) & 103.0 (14.7) & 103.2 (15.0) \\
			& Observations taken & 1002.5 & 1029.6 & 1032.5 & 1034.8 \\
			& Running Time (s) & 0.00 & 0.09 & 0.01 & 6.42\\
			\hline
			\multirow{4}{*}{(s2, 1, 0.5, 0.2)} & Coverage Probability (\%)	  & 93.8 & 94.6 & 94.4 & 92.6 \\
			& Half-width Achieved (\%) & 44.7 & 100 & 100 & 100 \\
			& Cost Gap \% (std.) & 100 (--) & 102.3 (11.4) & 102.5 (11.7) & 102.5 (11.6) \\
			& Observations taken & 1117.0 & 1142.3 & 1144.8 & 1145.1 \\
			\hline
			\multirow{4}{*}{(s3, 1, 0.5, 0.5)} & Coverage Probability (\%)	  & 93.7 & 94.5 & 96.7 & 95.4 \\
			& Half-width Achieved (\%) & 11.3 & 100 & 100 & 100 \\
			& Cost Gap \% (std.) & 100 (--) & 101.7 (2.2) & 102.3 (2.2) & 102.3 (2.2) \\
			& Observations taken & 1407.7 & 1435.2 & 1439.7 & 1439.8 \\
			\hline
			\hline
			\multirow{5}{*}{(s4, $\frac{1}{3}$, 0.3, 0.2)} & Coverage Probability (\%)	  & 92.3 & 94.4 & 94.5 & 94.6 \\
			& Half-width Achieved (\%) & 45.1 & 100 & 100 & 100 \\
			& Cost Gap \% (std.) & 100 (--) & 102.4 (18.2) & 102.9 (18.3) & 102.2 (18.2) \\
			& Observations taken & 1087.0 & 1114.6 & 1119.6 & 1101.4 \\
			& Running Time (s) & 0.00 & 0.10 & 0.01 & 6.84\\
			\hline
			\multirow{4}{*}{(s5, $\frac{1}{3}$, 0.5, 0.2)} & Coverage Probability (\%)	  & 94.5 & 95.1 & 95.6 & 93.2 \\
			& Half-width Achieved (\%) & 46.3 & 100 & 100 & 100 \\
			& Cost Gap \% (std.) & 100 (--) & 102.1 (16.2) & 102.3 (16.4) & 102.2 (16.2) \\
			& Observations taken & 1209.7 & 1235.7 & 1237.8 & 1265.8 \\
			\hline
			\multirow{4}{*}{(s6, $\frac{1}{3}$, 0.5, 0.5)} & Coverage Probability (\%)	  & 94.7 & 94.4 & 94.5 & 94.6 \\
			& Half-width Achieved (\%) & 9.8 & 100 & 100 & 100 \\
			& Cost Gap \% (std.) & 100 (--) & 101.9 (2.3) & 102.3 (2.4) & 102.2 (2.4) \\
			& Observations taken & 1524.1 & 1553.5 & 1559.1 & 1572.0 \\
			\hline
			\hline
			\multirow{5}{*}{(s7, 5, 0.3, 0.2)} & Coverage Probability (\%)	  & 93.5 & 95.1 & 94.5 & 95.4 \\
			& Half-width Achieved (\%) & 43.3 & 100 & 100 & 100 \\
			& Cost Gap \% (std.) & 100 (--) & 102.6 (13.4) & 103.0 (13.1) & 103.0 (13.5) \\
			& Observations taken & 1189.1 & 1221.0 & 1223.8 & 1203.8 \\
			& Running Time (s) & 0.00 & 0.11 & 0.01 & 7.54\\
			\hline
			\multirow{4}{*}{(s8, 5, 0.5, 0.2)} & Coverage Probability (\%)	  & 93.7 & 94.5 & 94.8 & 95.7 \\
			& Half-width Achieved (\%) & 39.6 & 100 & 100 & 100 \\
			& Cost Gap \% (std.) & 100 (--) & 102.3 (7.0) & 102.6 (7.0) & 102.4 (7.0) \\
			& Observations taken & 1321.6 & 1352.6 & 1356.3 & 1333.9 \\
			\hline
			\multirow{4}{*}{(s9, 5, 0.5, 0.5)} & Coverage Probability (\%)	  & 94.1 & 95.9 & 95.4 & 94.5 \\
			& Half-width Achieved (\%) & 11.2 & 100 & 100 & 100 \\
			& Cost Gap \% (std.) & 100 (--) & 102.0 (2.4) & 102.1 (2.4) & 102.2 (2.4) \\
			& Observations taken & 1664.8 & 1697.6 & 1699.7 & 1734.0 \\
			\hline
			\end{tabular}
		\caption{Comparison of results for the proposed cost-minimization procedures}
		\label{table:ResultsCostComparison1}
}
\end{table}

\subsection{Case Study}\label{sec:studycase}
 
According to the U.S. Food and Drug Administration (FDA), a generic drug is a copy of a brand-name drug; and both versions must have the same ``active ingredients, effectiveness, quality, safety, strength and benefits'', but they may look different \citep{FDA1}. Generic drugs (just as their brand-name counterparts) are approved by the FDA, and they are extremely important, as they allow  customer demand to be met, while keeping prices and insurance premiums low. In fact, generics represent 90\% of the total prescriptions made in the U.S. \citep{FDA2} and saved the health care system 338 billion dollars in 2020, with a 10-year projected savings of \$2.4 trillion \citep{AAM1}. 

Generic drugs do not need to pass extensive clinical trials as did their brand-name counterparts.  Instead, generic applicants only need to prove that their drug is bioequivalent to the respective brand-name drug \citep{FDA3}, which is accomplished by comparing the peak drug concentration and area under the curve (describing the rate and extent of absorption of the drug) vs.\ the same measures for the respective brand-name drug. The ratio of these measures must lie between 0.80 and 1.25; a more-detailed explanation can be found in \cite{Andrade2015}. 

A potential issue with this methodology is that patient outcomes are not measured directly, and bioequivalent medicines may actually lead to different health outcomes. \cite{Davit2009} presented a review of 12 years worth of data, showing that generic and brand-name products did not significantly differ when compared based on the FDA's bioequivalence criteria; however, several studies have reported differences in the efficacy (health outcomes) of generic drugs as compared to their corresponding brand-name versions \citep{Galleli2013}. Because of the potential differences in health outcomes, many researchers around the world have focused their efforts on identifying such scenarios, as evidenced by \cite{Leclerc2017}, \cite{Desai2019}, and \cite{Tian2020}. In particular, \cite{Leclerc2017} presented confidence intervals for the difference of the rate of adverse events (generic vs.\ corresponding brand-name drug) for three generic drugs (Losartan, Valsartan, and Candesartan). The mean increases for the three generic drugs are 8.0\%, 11.7\%, and 16.6\%, respectively; and an analysis of the study, plus economic and societal implications of such findings, was conducted by \cite{Alter2017}. 

Given the previous context, our methodology could be used to improve the understanding regarding the difference of probabilities of bad outcomes, $p_x-p_y$, between generic and brand-name drugs, all while minimizing the total costs of the sampling (including costs related  to the number of people who need to be involved in the test). This has huge implications for society and also may help decision makers to decide on whether a generic drug should be approved or not (evaluating the trade-offs between savings in drug cost, but also in potential worse outcomes in case the generic drug does not work as well as the brand-name counterpart). 

To illustrate the benefits of our work, we consider the generic drug Valsartan (denoted V) vs.\ its brand-name counterpart Diovan (denoted D)\@. For purposes of illustration, we assume that the bad-outcome probabilities are $p_{D}=0.1$ and $p_{V}=p_{D}+0.117=0.217$. The drugs' prices range from \$259 to \$280 per month (for Diovan) and \$14 to \$38 per month (for Valsartan) \citep{Prices2022}. We will test three different configurations for the prices: $(c_{D}=259,c_{V}=14),\ (c_{D}=280,c_{V}=38)$, and $(c_{D}=259,c_{V}=14)$. For simplicity, we assume that outcomes can be measured after one month of treatment. We seek a 95\% CI for the difference $p_D-p_V$, and will compute results for $\epsilon=0.02$ (thus returning a CI of total width of at most 4\%) and $\epsilon=0.015$ (total width at most 3\%). In addition, we assume that 500 people can participate in the study every month ($B=500$).  We will test the performance of our Algorithm \ref{Algo:BatchSeqMinCost} under two settings: the minimization of costs, and the minimization of observations (the latter by naively considering the same cost for both drugs). Our baseline is the sequential Algorithm \ref{Algo:NaiveSeqMinCost}. Results for 10,000 replications for each scenario are given in Tables \ref{table:studycase1} and \ref{table:studycase2}.

\begin{table}[!h]
		\centering
		\begin{tabular}{|c|c|c|c|c|}
			\hline
				$(c_{D},c_{V})$  &  Performance Measures   & Algorithm \ref{Algo:NaiveSeqMinCost} & Algorithm \ref{Algo:BatchSeqMinCost} & Algorithm \ref{Algo:BatchSeqMinCost}    \\
					 &  & Baseline\ & min.\ observations &  min.\ costs   \\
			\hline \hline
			\multirow{4}{*}{(259, 14)} & Coverage Probability (\%)	  & 94.9 & 94.9 & 94.6 \\
			& Average Costs (\%) & 714,550 & 526,937 & 407,385  \\
			& Months Required for Sampling &[9,11] & [9,11] & [13,19] \\
			& Average Sampling Time (Mo) & 11 & 11 & 17 \\
			\hline
			\multirow{4}{*}{(259, 38)} & Coverage Probability (\%)	  & 94.9 & 94.6 & 94.6 \\
			& Average Costs (\%) & 777,368 & 591,541 & 545,252  \\
			& Months Required for Sampling &[9,11] & [9,11] & [11,14] \\
			& Average Sampling Time (Mo) & 11 & 11 & 13 \\
			\hline
            \multirow{4}{*}{(280, 38)} & Coverage Probability (\%)	  & 94.9 & 94.6 & 94.6 \\
			& Average Costs (\%) & 832,333 & 631,210 & 573,097  \\
			& Months Required for Sampling &[9,11] & [9,11] & [13,19] \\
			& Average Sampling Time (Mo) & 11 & 11 & 17 \\
			\hline
			\hline
		    \end{tabular}
		\caption{Case study results for $\epsilon=0.02$}
		\label{table:studycase1}
	\end{table}

\begin{table}[!h]
		\centering
		\begin{tabular}{|c|c|c|c|c|}
			\hline
				$(c_{D},c_{V})$  &  Performance Measures   & Algorithm \ref{Algo:NaiveSeqMinCost} & Algorithm \ref{Algo:BatchSeqMinCost} & Algorithm \ref{Algo:BatchSeqMinCost}    \\
					 &  & Baseline\ & min.\ observations &  min.\ costs   \\
			\hline \hline
			\multirow{4}{*}{(259, 14)} & Coverage Probability (\%)	  & 94.6 & 94.9 & 94.6 \\
			& Average Costs (\%) & 1,243,017 & 975,939 & 710,368  \\
			& Months Required for Sampling &[17,19] & [16,19] & [25,32] \\
			& Average Sampling Time (Mo) & 19 & 18 & 29 \\
			\hline
			\multirow{4}{*}{(259, 38)} & Coverage Probability (\%)	  & 94.6 & 94.7 & 94.6 \\
			& Average Costs (\%) & 1,352,293 & 1,093,770 & 950,534  \\
			& Months Required for Sampling &[17,19] & [16,19] & [20,24] \\
			& Average Sampling Time (Mo) & 19 & 18 & 22 \\
			\hline
            \multirow{4}{*}{(280, 38)} & Coverage Probability (\%)	  & 94.9 & 94.7 & 94.6 \\
			& Average Costs (\%) & 1,447,910 & 1,167,327 & 1,001,178  \\
			& Months Required for Sampling &[17,19] & [17,19] & [20,25] \\
			& Average Sampling Time (Mo) & 19 & 18 & 23 \\
			\hline
			\hline
		    \end{tabular}
		\caption{Case study  results for $\epsilon=0.015$}
		\label{table:studycase2}
	\end{table}

A smaller value of $\epsilon$ obviously requires more observations, and so more sampling stages. This is clearly apparent from Tables \ref{table:studycase1} and \ref{table:studycase2}, as the number of sampling stages (i.e., months) increases as well as the cost. From Table \ref{table:studycase1}, we see that when we use Algorithm \ref{Algo:BatchSeqMinCost}, the coverage seems to drop by just a small amount, but there is a significant decrease with respect to costs, regardless of whether the algorithm is used to minimize the number of observations or if it is used to minimize the costs; in fact, in each of the 10,000 replications for each scenario, the cost turned out to be smaller than the cost obtained when using the baseline algorithm. We see that when using Algorithm \ref{Algo:BatchSeqMinCost} to minimize the number of observations, there is no deduction in the time required for the study to be performed (i.e., the number if stages), and the cost savings are around \$180--200k. When using Algorithm \ref{Algo:BatchSeqMinCost} to minimize costs, there is an increase in the time the study needs to be performed (between 2--6 months), but savings increase, and are in the \$230--307k interval.

When comparing the performance measures in Table \ref{table:studycase2}, since we now require more sampling epochs (i.e., months), the differences in coverage probability of our two methods vs.\ the baseline are insignificant. The use of Algorithm \ref{Algo:BatchSeqMinCost} to minimize observations starts to return a reduction on the sampling time (before there was a decrease on the average number of observations, but not the average sampling time because of discretization over months), while also continuing to have savings on every realization of the sampling, with average savings between \$260--280k dollars. When using Algorithm \ref{Algo:BatchSeqMinCost} to minimize costs, the increase in sampling time is between 3 and 10 months, and savings are in the \$400--530k range. It is also clear that more savings are possible when the proportion $c_{V}/c_{D}$ is smaller; however, that also increases the difference in sampling time for Algorithm \ref{Algo:BatchSeqMinCost} when minimizing for costs. This suggests that when that value is very small, then minimizing observations with that algorithm is the best alternative as the performance on sampling epochs (compared to the baseline) is not affected; but when that ratio increases, more economic savings can be obtained by using the algorithm to minimize costs, and with a small trade-off in sampling epochs.

As a recapitulation, it is easy to see how our algorithms provide a significant value for such applications as testing over the health outcomes for generic and brand-name drugs --- and doing so while saving significant costs carries benefits for society.

\section{Summary and Conclusions}
\label{sec:conclusions}

This paper is concerned with the efficient computation of confidence intervals for the difference between parameters from two Bernoulli populations, where efficiency is defined in terms of sampling costs. As such, we proposed four different procedures to minimize the total cost of the sampling, depending on the type of sampling scheme. One procedure is undertaken in two stages, whereas the others carry out several stages and adjust their policies according to the real-time sampling results. We showed that our two-stage cost-minimization procedure (Algorithm \ref{Algo:2StageMinCost}) can result in savings of up to 50\% vs.\ the conservative approach depicted in Algorithm \ref{Algo:Conservative} and up to 14\% vs.\ the counterpart two-stage procedure that minimizes the number of observations taken (Algorithm \ref{Algo:2StageMinSamples}). Furthermore, the proposed sequential methods save up to 15\% in costs compared to the baseline ``naive'' algorithm (Algorithm \ref{Algo:NaiveSeqMinCost}) and up to 14\% vs.\ their counterparts minimizing the number of total observations. In addition to the cost savings, our proposed batching-based sampling scheme (Algorithm \ref{Algo:BatchSeqMinCost}) always obtains the desired half-width, while approximately attaining the specified coverage. Further, the batching scheme reduces the number of decision epochs (compared to fully sequential methods) at the small price of just an extra 2--3\% of costs vs.\ the two-stage cost-saving procedure (that is {\em not guaranteed\/} to actually yield the desired half-width). Our case study on generic vs.\ brand-name drug efficacies illustrated the cost-effectiveness potential of our new methodology, where significant cost savings were obtained by our new algorithms on a realistic and useful application.

There are a large number of interesting problems that have not been addressed here. First, our one-step look-ahead methodology produces similar results, but is much more computationally expensive, so it makes sense to use the batching approach; this future research will be undertaken in order to study the potential advantage of adding more steps in the relevant expected value computations. Second, additional research may also include testing the performance of our algorithms on borderline scenarios (cases in which $p_x$ and/or $p_y$ are very close to 0 or 1, e.g., rare-event simulation) or considering a guarantee on the coverage probability (for example, worst-case analysis or a weighted average across a large set of scenarios).  
Third, it would certainly be of interest to adopt a Bayesian approach \citep[see, for instance,][]{lehmann98}
that might allow us to find a procedure that can balance the tradeoff between three criteria: expected cost, length of half-width, and coverage probabilities over the entire range of $(p_x,p_y)$ values. Thus, the current work should be considered as a starting point for several different future investigation streams.




\appendix
\section{Benchmarking Algorithms from \cite{erazo21}}\label{App}

This appendix presents the implementations used by several  algorithms from \cite{erazo21} to benchmark the algorithms proposed in this paper.


\begin{algorithm}
\caption{Two-Stage Procedure to minimize the number of observations taken}\label{Algo:2StageMinSamples}
\textbf{Notation:} $m_{x,2}, m_{y,2}$: total numbers of observations to take from populations $X$ and  $Y$, respectively. 
\begin{algorithmic}[1]
\Require initial sample sizes $m_{x,1},  m_{y,1} > 0$, initial sample means $\bar{X}_1, \bar{Y}_1$, desired coverage probability $(1-\alpha)$, and desired half-width $\epsilon$
\State $\tau_{x,1} \gets \bar{X}_1(1-\bar{X}_1)$, \,  $\tau_{y,1} \gets \bar{Y}_1(1-\bar{Y}_1)$
\State $\beta^\star \gets \dfrac{1}{1+\sqrt{ \tau_{y,1} /\tau_{x,1} }}$
\State $M^\star \gets \dfrac{z_{\alpha/2}^2 (\sqrt{\tau_{x,1}}+\sqrt{\tau_{y,1}})^2 }{\epsilon^2}$
\State $m_{x,2} \gets \max(\lfloor \beta^\star M^\star \rceil , m_{x,1} )$, \,  $m_{y,2} \gets \max(\lfloor (1-\beta^\star) M^\star \rceil , m_{y,1} )$
\State Sample $m_{x,2} - m_{x,1}$ observations from population $X$ and $m_{y,2} - m_{y,1}$ from population $Y$
\State Compute overall sample means $\bar{X}_2$ and $\bar{Y}_2$
\Ensure Confidence Interval $\bar{X}_2 - \bar{Y}_2 \pm z_{\alpha /2} \sqrt{ \frac{\bar{X}_2(1-\bar{X}_2)}{m_{x,2}} + \frac{\bar{Y}_2(1-\bar{Y}_2)}{m_{y,2}} }$
\end{algorithmic}
\end{algorithm}

\newpage

\begin{algorithm}
\caption{Fully Sequential Procedure to minimize the number of observations taken}\label{Algo:FullySeqMinSamples}
\textbf{Notation:} $m_{x,r},  m_{y,r}$: cumulative numbers of observations taken from populations $X$ and $Y$ by the end of stage $r$, respectively. $\bar{X}_r, \bar{Y}_r$: sample means at the completion of stage $r$ for populations $X$ and $Y$.
\begin{algorithmic}[1]
\Require initial sample sizes $m_{x,1},  m_{y,1}$, initial sample means $\bar{X}_1, \bar{Y}_1$, desired coverage probability $(1-\alpha)$, and desired half-width $\epsilon$
\State $r \gets 2$, Continue $\gets$ $True$
\While{Continue == $True$}
    \State $\tau_{x,r-1} = \bar{X}_{r-1}(1-\bar{X}_{r-1})$, \ $\tau_{y,r-1} = \bar{Y}_{r-1}(1-\bar{Y}_{r-1})$
    \If{$\tau_{x,r-1} \left[ \frac{1}{m_{x,r-1}} - \frac{1}{m_{x,r-1}+1} \right] \; > \;
\tau_{y,r-1} \left[ \frac{1}{m_{y,r-1}} - \frac{1}{m_{y,r-1}+1} \right]$}
        \State Sample from population $X$
        \State $m_{x,r}\gets m_{x,r-1}+1$, \ $m_{y,r}\gets m_{y,r-1}$
        \State Compute $\bar{X}_r$ and $\bar{Y}_r\gets \bar{Y}_{r-1}$
    \Else
        \State Sample from population $Y$
        \State $m_{x,r}\gets m_{x,r-1}$, \ $m_{y,r}\gets m_{y,r-1}+1$
        \State $\bar{X}_r\gets \bar{X}_{r-1}$ and  compute $\bar{Y}_r$
    \EndIf
    \State $H_r\gets z_{\alpha/2}\sqrt{ \dfrac{\bar{X}_r (1-\bar{X}_r)}{m_{x,r}} + \dfrac{\bar{Y}_r (1-\bar{Y}_r)}{m_{y,r}}}$
    \If{$H_r\leq \epsilon$}
        \State Continue $\gets$ $False$
    \Else
        \State $r \gets r+1$
    \EndIf
\EndWhile
\Ensure Confidence Interval $\bar{X}_r - \bar{Y}_r \pm H_r$
\end{algorithmic}
\end{algorithm}

\newpage

\begin{algorithm}
\caption{Sequential Procedure with Batching to minimize the number of observations taken}\label{Algo:BatchSeqMinSamples}
\textbf{Notation:} $m_{x,r},  m_{y,r}$: cumulative numbers of observations taken from populations $X$ and $Y$ by the end of stage $r$, respectively. $\bar{X}_r, \bar{Y}_r$: sample means at the completion of stage $r$ for populations $X$ and $Y$.
\begin{algorithmic}[1]
\Require initial sample sizes $m_{x,1}, m_{y,1}$, initial sample means $\bar{X}_1, \bar{Y}_1$, batch size $B$, desired coverage probability $(1-\alpha)$, and desired half-width $\epsilon$
\State $\tau_{x,1} \gets \bar{X}_1(1-\bar{X}_1) $, \
 $\tau_{y,1} \gets \bar{Y}_1(1-\bar{Y}_1) $
\State $r \gets 2$, Continue $\gets$ $True$
\While{Continue == $True$}
    \State $\gamma_r^\star \; \gets \; \dfrac{\sqrt{\tau_{x,r-1}}\,(m_{y,r-1}+B)-\sqrt{\tau_{y,r-1}}\,m_{x,r-1}}{B\big(\sqrt{\tau_{x,r-1}}+\sqrt{\tau_{y,r-1}}\,\big)}$
    \If{$\gamma_r^\star < 0$}
        \State $\gamma_r^\star \gets 0$
    \EndIf
    \If{$\gamma_r^\star > 0$}
        \State $\gamma_r^\star \gets 1$
    \EndIf
    \State Sample $\lfloor \gamma_r^\star B\rceil$ observations from population $X$ and $\lfloor (1-\gamma_r^\star) B\rceil$ observations from $Y$
    \State $m_{x,r}\gets m_{x,r-1}+\lfloor \gamma_r^\star B\rceil$, \
  $m_{y,r}\gets m_{y,r-1}+\lfloor (1-\gamma_r^\star) B\rceil$
    \State Compute sample means $\bar{X}_r$ and $\bar{Y}_r$
    \State $\tau_{x,r}\gets \bar{X}_r(1-\bar{X}_r)$, \
    $\tau_{y,r}\gets \bar{Y}_r(1-\bar{Y}_r)$

    \State $H_r\gets z_{\alpha/2}\sqrt{ \dfrac{\tau_{x,r}}{m_{x,r}} + \dfrac{\tau_{y,r}}{m_{y,r}}}$
    \If{$H_r\leq \epsilon$}
        \State Continue $\gets$ $False$
    \Else
        \State $r \gets r+1$
    \EndIf
\EndWhile
\Ensure Confidence Interval $\bar{X}_r - \bar{Y}_r \pm H_r$
\end{algorithmic}
\end{algorithm}

\section{Intractable Markov Decision Process}
\label{sec:MDP}

We now discuss an optimal finite-horizon Markov decision process formulation
that minimizes 
the expected cost of obtaining the CI having the desired half-width. As in \S\ref{sec:SS-group}, we assume that there is a maximum sampling capacity of batch size $B$ per period. We also introduce the following standard MDP notation and relate this in the context of our fixed-width CI problem.
\begin{itemize}
    \item ${\cal T} = \{1,2,\ldots,T\}$ is the {\em time horizon\/}, where $T$ is a very large number that is an upper bound on the number of sampling epochs that will be required.
    
    \item   $\mathcal{S}=\{(w_x,m_x,w_y,m_y) :  0 < w_x < m_x\leq BT, \, 0 < w_y < m_y\leq BT \}$ is the {\em state space\/}, where $w_x, w_y$ represent the numbers of successes, and $m_x, m_y$ represent the numbers of observations taken (so far) from populations $X$ and $Y$.
    
    \item $A_s$ is the {\em action space\/} for $s = (w_x,m_x,w_y,m_y)\in \mathcal{S}$, where
 
     $A_{s}= \begin{cases}(0,0), \ \text{if } z_{\alpha/2}\sqrt{ \frac{ \hat{p}_x(1-\hat{p}_x) }{ m_x } + \frac{ \hat{p}_y(1-\hat{p}_y) }{ m_y } } \leq \epsilon,  \text{ for } \hat{p}_x=w_x/m_x \text{ and }  \hat{p}_y= w_y/m_y;  \\ (b_x,b_y) :  b_x+b_y = B, \ \text{otherwise.} \end{cases}$ 
    
   In other words, no actions are taken if the width of the CI has been achieved; otherwise, we take $b_x, b_y$ observations (a total of $B$) from populations $X$ and $Y$, respectively, in the actual decision epoch.
    
    \item For current state $s = (w_x,m_x,w_y,m_y) \in {\cal S}$ and action $a = (b_x,b_y)\in {\cal A}_s$, we let $P_{s,a}$ denote a $(b_x+1) \times (b_y+1)$ {\em probability matrix\/} having  $(k_x,k_y)$th element 
    \begin{eqnarray*}
    P_{s,a}(k_x,k_y) & = & P\big( w_x+k_x,m_x+b_x,w_y+k_y,m_y+b_y \big| s,a  \big)  \\
    & = &  { b_x \choose k_x } \hat{p}_x^{k_x} (1-\hat{p}_x)^{b_x-k_x}  { b_y \choose k_y } \hat{p}_y^{k_y} (1-\hat{p}_y)^{b_y-k_y},
    \end{eqnarray*}
    for all $0\leq k_x\leq b_x , \ 0 \leq k_y \leq b_y$; this is simply the product of two binomial probabilities, calculated by assuming $\hat{p}_x$ and $\hat{p}_y$ are the true success probabilities and using the fact that all observations from both populations are independent. It follows that, given action $a$,  the transition probabilities from state $s$ into state $s' = (w_x',m_x',w_y',m_y')$ are given by:
    
    $P(s'|s,a)= \begin{cases} P_{s,a}(w_x'-w_x,w_y'-w_y) , \ \text{for } w_x\leq w_x'\leq w_x+b_x , \ w_y\leq w_y'\leq w_y+b_y \\ \hspace{4cm} \text{ and } m_x+b_x=m_x' , \ m_{y}+b_{y}=m_{y}'   \\ 0, \ \text{otherwise.} \end{cases}$ 

    \item The terminal rewards $r_T(w_x,m_x,w_y,m_y) = - (m_xc_x+m_yc_y)$ represent the cost of the sampling procedure at the last possible decision epoch $T$, whereas all the other (intermediate) rewards are equal to zero.
\end{itemize}

By maximizing the expected rewards, we minimize the sampling costs and obtain a policy that tells us (i) how to distribute the $B$ observations in each active sampling day/period, and (ii)  when to stop sampling.  This is a sequential decision/sampling method --- at each decision epoch, the corresponding optimal decision minimizing the expected costs can be determined given the actual realization of the system. We note that this method does not necessarily require an initial sample, as the minimax estimators \citep{Sh21} can be used for $\hat{p}_x$ and $\hat{p}_y$ to guarantee that both will be in the interval $(0,1)$  --- the use of minimax estimators would change the definition of the probability matrix slightly, where the state space ${\cal S}$ would now be $0\leq w_x\leq m_x \leq BT$, $0\leq w_y \leq m_y \leq BT$. In any case, the constraint on the action space $A_s$ can be relaxed to $b_x+b_y\leq B$, which would make sense if only a few observations are needed towards the final decision epochs (perhaps because the actual interval is very close to the desired length).

The issue with this formulation is that it is likely to be intractable, as $BT$ is a potentially large upper bound on the number of observations, and $|\mathcal{S}| = O(B^4T^4)$. This renders as problematic the computation of the MDP's recursion to obtain the optimal policy; and thus, in practice, we cannot solve this MDP or its dual Linear Programming (LP) version. Our $\ell$-stage algorithm is a relaxation of the MDP as it does not consider the expectation cost over all the next stages, but when $\ell$ goes to infinity, we recover the same exact structure (as we will never use the ``cost-to-go'' function if enough stages are considered).

\end{document}